\documentclass[12pt]{article}
\usepackage{ifpdf}
\pdfoutput=1
\usepackage[a4paper,width=18cm, bottom=3.2cm, left=1.5cm]{geometry}
\usepackage{fancyhdr}
\usepackage{lipsum}
\usepackage[utf8x]{inputenc}
\usepackage{graphicx,caption}
\usepackage{hyperref}
\usepackage[font=small,labelfont=bf]{caption}

\usepackage[utf8x]{inputenc}
\usepackage{amssymb}
\usepackage{amsmath}
\usepackage{graphicx}
\usepackage{hyperref}
\usepackage{color}
\usepackage[font=small,labelfont=bf]{caption}
\newcommand{\nn}{\nonumber}

\newcommand{\be}{\begin{equation}}
\newcommand{\ee}{\end{equation}}
\newcommand{\bea}{\begin{eqnarray}}
\newcommand{\eea}{\end{eqnarray}}

\newcommand{\dd}{ \textmd{d} }
\newcommand{\rf}[1]{(\ref{#1})}
\newcommand{\beal}[1]{\begin{eqnarray}\label{#1}}
\newcommand{\eeal}{\end{eqnarray}}
\newcommand{\bel}[1]{\begin{equation}\label{#1}}
\newcommand{\eel}{\end{equation}}

\newcommand{\f}[2]{\frac{#1}{#2}}

\usepackage{microtype}

\usepackage{authblk}

\title{\bf Quasinormal modes and the phase structure of strongly coupled matter }

\author{Romuald A. Janik
\footnote{Email:romuald@th.if.uj.edu.pl}}
\author{Jakub Jankowski
\footnote{Email:jakubj@th.if.uj.edu.pl}}
\author{Hesam Soltanpanahi
\footnote{Email:hesam@th.if.uj.edu.pl}}
\affil{Institute of Physics, Jagiellonian University, {\L}ojasiewicza 11, 30-348  Krak{\'o}w, Poland}
\date{}

\begin{document}

\maketitle

\thispagestyle{empty}

\abstract{ We investigate the poles of the retarded Green's 
functions of strongly coupled
field theories exhibiting a variety of phase structures
from a crossover up to different first order phase transitions.
These theories are modeled by a dual gravitational description. 
The poles of the holographic Green's functions appear at the
frequencies of the quasinormal modes of the dual black hole background.
We focus on quantifying linearized level dynamical response of the system
in the critical region of phase diagram.  Generically non-hydrodynamic degrees
of freedom are important for the low energy physics in the vicinity of a phase transition. 
For a model with linear confinement
in the meson spectrum we find
degeneracy of hydrodynamic
and non-hydrodynamic modes close to the minimal black hole temperature, and we establish a region of temperatures with
unstable non-hydrodynamic modes in a branch of black hole solutions.}

\newpage
\tableofcontents


\section{Introduction}

It is almost twenty years since there has been discovered a remarkable new relation between 
geometry and physics: within the Anti-de Sitter/Conformal Field Theory (AdS/CFT) 
correspondence  \cite{Maldacena:1997re} we can investigate the dynamics of 
strongly coupled quantum field theories by means of General Relativity methods. From purely academic studies this 
field of research evolved to address experimental systems an 
example being strongly interacting hadronic matter \cite{CasalderreySolana:2011us}.
In particular, real time response of a thermal equilibrium state has been quantified in the case of $\mathcal{N}=4$
super Yang-Mills theory by the means of the poles of the retarded Green's function \cite{Kovtun:2005ev},
which correspond to quasinormal modes (QNM) 
in the dual gravitational 
theory.

While the hydrodynamic QNMs have been studied in different gravitational theories dual to non-CFT cases (e.g. ref. \cite{Benincasa:2005iv,Gursoy:2013zxa}),
initial steps towards extension  were taken in ref. \cite{Janik:2015waa,Buchel:2015saa}
where nonhydrodynamic QNM's of an external scalar field were considered in non-conformal field theories, which 
still admit a gravitational dual description. Subsequent investigations include different mechanisms of scale 
generation \cite{Buchel:2015ofa}, different relaxation channels \cite{Attems:2016ugt,Ali-Akbari:2016sms},
baryon rich plasma \cite{Rougemont:2015wca}, and studies of non-relativistic systems \cite{Gursoy:2016tgf}.

This paper is an extended version of the letter \cite{Janik:2015iry} where we provide many more details as
well as extend the investigation to a model of an improved holographic QCD type which exhibits
novel and interesting phenomena. We concentrate on investigating linearized real time response
of strongly coupled non-conformal field theories in the vicinity of various types of phase transitions
and phase structures. Thus the physical regime of interest in the present paper is quite distinct from the 
the one of interest for `early thermalization' which have been extensively studied within the
AdS/CFT correspondence.

Firstly, we analyze all allowed channels of energy-momentum tensor perturbations and corresponding 
two-point correlation functions. Secondly, we concentrate on the phenomena appearing in the vicinity
of a nontrivial phase structure of various type: a crossover (motivated by the lattice QCD equations of state \cite{Borsanyi:2012cr}), a $2^{\rm nd}$ order phase transition and a $1^{\rm st}$ order  
phase transition. These cases are modeled by choosing appropriate scalar field self-interaction potentials
in a holographic gravity-scalar theory used in \cite{Gubser:2008ny}.
Apart form this, we also analyze a potential from a different family of models, improved holographic QCD (IHQCD), considered in \cite{Gursoy:2008za,Gursoy:2008bu}. In this case the focus was 
on getting best possible contact with properties of QCD, in particular 
asymptotic freedom and colour confinement as well as obtaining a realistic value
of the bulk viscosity.

Despite the fact, that considered models have a rather simplistic construction, the resulting near equilibrium
response shows a variety of non-trivial phenomena. Some generic features consist of:
\emph{(i)} the breakdown  of the applicability
of a hydrodynamic description already at lower momenta than in the conformal
case; 
\emph{(ii)} in the cases with a first order phase transition we find a generic minimal
temperature, $T_m$, below which no unstable solution exists; \emph{(iii)} whenever there exists a thermodynamical instability there is a corresponding
dynamical instability present in the hydrodynamic mode of the theory; 
\emph{(iv)} the \emph{ultralocality} property of non-hydrodynamic modes, i.e., weak dependence
on the momentum scale. 

The nature of the dual gravitational formulation allows for a detailed quantitative investigation of the above 
phenomena as well as for accessing diverse physical scenarios.
In particular, the first order phase transition appears in two different scenarios. The first one is 
similar to the usual Hawking-Page transition \cite{Hawking:1982dh} in which the two phases are 
a black hole geometry and a thermal gas geometry \cite{Gursoy:2008bu}. In the second one the transition appears between two black hole solutions \cite{Gubser:2008ny}. 
This diversity is triggered by a different functional dependence of the scalar field potential in the 
deep infrared (IR) region, and is reflected in the corresponding QNM spectrum. Nevertheless there is a common aspect in both situations. We observe some specific dynamical response of the system for a \emph{characteristic} temperature, $T_{\rm ch}>T_m$, in the stable branch of EoS. The details of 
this effect depend on the case, but the existence of $T_{\rm ch}$ is generic for a first order
phase transition. 

Particularly interesting  effects appear in IHQCD model, which admits a first order 
phase transition between a black hole and a thermal gas \cite{Gursoy:2008bu}.
First, for temperatures in the range $T_m\leq T\leq T_{\rm ch}$ the lowest lying excitation modes become purely imaginary for low momenta, which leads to a ultralocality violation. 
Second, at $T=T_m$ for momenta higher than some threshold value the hydrodynamic mode and
the first non-hydrodynamic mode have the same dispersion relation.
Third, in the small black hole branch there is a range of temperatures which shows instability in a non-hydro mode.
 The appearance of these
phenomena makes the IHQCD model unique in the landscape considered.

The organization of the paper is as follows. In the next section, \ref{Background} 
we 
shortly describe the thermodynamics of considered models and parameter
choices for bulk scalar interactions. In section \ref{QNM} we discuss
equations of motion for the linear perturbations of the background and 
technical aspects of their solutions. In the first subsection we clarify the 
right boundary conditions which have to be chosen for the QNM spectrum.
In the second subsection we give general remarks and list main aspects of physical 
properties we obtain.
The following sections \ref{secVQCD} to \ref{ihqcd} contain results and detailed studies of different cases.
We close the paper by a summary and outlook in section \ref{summary}. 
For completeness appendixes \ref{AppB} and \ref{app}  respectively contain some technical details of the Free Energy computation, and the explicit form of the 
QNM equations of motion.


\section{The background and thermodynamics of the system}
\label{Background}

In this section we formulate the background black hole solutions and
determine the scalar field potential by considering emergent equations of
state in the dual field theory.


\subsection{Metric Ansatz and equations of motion}

This section describes the black hole background solutions for the quasinormal
mode calculations, which follow from the action
\be
S=\frac{1}{2\kappa_5^2}\int_{\mathcal{M}} d^5x \sqrt{-g}  \left[ R-\f{1}{2}\, \left( \partial \phi \right)^2 - V(\phi) \, \right]
-
\frac{1}{\kappa_5^2}\int_{\partial\mathcal{M}}d^4x \sqrt{-h}~ K~,
\label{Action}
\ee 
where $V(\phi)$ is thus far arbitrary and $\kappa_5$ is related to  five dimensional 
Newton constant by $\kappa_5=\sqrt{8\pi G_5}$.
The last term in \eqref{Action} is the standard Gibbons-Hawking boundary contribution.
These solutions are similar to those studied in ref. \cite{Gubser:2008ny,Gursoy:2008za}.
Since our goal is to determine the QNM  frequencies, it will 
be convenient to employ Eddington-Finkelstein coordinates, which 
have been proven useful in the case of the scalar field modes \cite{Janik:2015waa}. 
We will discuss this in a more detail in the following section.

Whereas we are interested in asymptotically AdS space-time geometry, 
the potential needs to have the following small $\phi$ expansion
\be
V(\phi)\sim -\frac{12}{L^2} + \frac{1}{2}m^2\phi^2 + O(\phi^4)~. 
\ee
Here, $L$ is the $AdS$ radius, which we set it to one, $L=1$, by the 
freedom of the choice of units.
Such a gravity dual corresponds to relevant deformations
of the boundary conformal field theory
\be
\mathcal{L} = \mathcal{L}_{\rm CFT} + \Lambda^{4-\Delta} O_{\phi}~,
\ee
where $\Lambda$ is an energy scale, and $\Delta$ is a conformal
dimension of the operator $O_\phi$ related to the mass parameter of the scalar field according to holography, $\Delta(\Delta-4)=m^2$. 
We consider $2\leq\Delta<4$ which corresponds to relevant 
deformations of the CFT and satisfies the Breitenlohner- Freedman bound, $m^2\geq-4$ \cite{Breitenlohner:1982bm,Breitenlohner:1982jf}.

The Ansatz for solutions under considerations
follows from the assumed symmetries: translation invariance in the
Minkowski directions as well as $SO(3)$ rotation symmetry in the spatial part. 
This leads to the following form of the line element:
\be
\label{linel0}
\dd s^2 = g_{tt} \dd t^2 + g_{xx} \dd \vec{x}^2 + g_{rr} 
\dd r^2 + 2 g_{rt} \dd r \dd t~,
\ee
where all the metric coefficients appearing in~\rf{linel0} are functions of the
radial coordinate $r$ 
alone, as is the scalar field $\phi$.
This form of the field Ansatz
(determined so far only by the assumed symmetries) 
allows two gauge 
choices to be made.
For the purpose of computing the quasinormal modes it is very convenient to
use the Eddington-Finkelstein gauge $g_{rr}=0$. It is
typically convenient also to impose the gauge choice  $g_{tr}=1$, but for our
purposes it turns out to be very effective to use the remaining gauge
freedom to set $\phi = r$. We label the metric components as 
\beal{linelef}
\dd s^2 &=&  e^{2 A} \, (-h \,\dd t^2 + \dd \vec{x}^2) - 2\, e^{A+B}\, \dd t\, \dd r~, \\
\phi &=& r~.
\eea
In the above coordinate system the UV boundary is at $r=0$, while the IR region
is the limit $r\rightarrow\infty$.
The system of Einstein-scalar field equations 
\begin{eqnarray}
&&R_{\mu\nu}-\frac{1}{2}\nabla_\mu\phi\nabla_\nu\phi-\frac{1}{3}V(\phi)g_{\mu\nu} = 0~,\\
&&\nabla_\mu\nabla^\mu\phi-\frac{d V(\phi)}{d\phi} = 0~,
\end{eqnarray}
takes the following form 
\begin{align}
\label{eq1}
&&A''-A'B'+\frac{1}{6}=0~,
\\[0ex]
\label{eq2}
&&h''+(4 A'-B')h'=0~,
\\[0ex]
\label{eq3}
&&6 A' h' + h(24 A'^2-1)+2 e^{2 B} V=0~,
\\[0ex]
\label{eq4}
&&4 A'-B'+\frac{h'}{h}-\frac{e^{2 B}}{h}V'=0~,
\end{align}
where the prime denotes a derivative with respect to $\phi$. 

In contrast to methods proposed in ref. \cite{Gubser:2008ny}
we solve this coupled equations directly using the spectral
method \cite{Grandclement:2007sb} in the Newton linearization
algorithm. 
We are interested in solutions possessing a horizon, which requires that the blackening
function $h(r)$ should have a zero at some $r=r_H$:
\bel{hzero}
h(r_H) = 0~.
\ee 
Asymptotically we require that our geometry is that of the $AdS$ space-time.


\subsection{Thermodynamics}
\label{secThermo}

Having determined the geometry we can extract the thermodynamic quantities 
in a standard way.
The Bekenstein-Hawking formula for entropy, together with the event
horizon regularity, lead to the following expressions
for the entropy density and the Hawking temperature
\bel{entrodens}
s = \f{2\pi}{\kappa_5^2} e^{3A(r_H)}~,
\hspace{35pt}
T = \f{e^{A(r_H) + B(r_H)} |V'(r_H)|}{4\pi}~. 
\ee
In turn, the speed of sound of the system can be determined as
\be
c_s^2 = \frac{d \log T}{d \log s}~.\label{Cs2}
\ee
Let us emphasis that this is the speed of sound in the 
dual field theory.
The corresponding Free Energy (FE) is related to the value of 
the action evaluated at the solution \cite{Witten:1998zw}
\begin{equation}
\mathcal{\beta F}= \lim_{\epsilon\rightarrow0}\left(S(\epsilon)-S_{\rm ct}(\epsilon)\right)~,
\label{eq:FE}
\end{equation}
where $\beta=1/T$, $S$ is the Einstein-Hilbert-scalar action (with Gibbons-Hawking term) evaluated on-shell with a cut-off
$\epsilon$ in a holographic direction. $S_{\rm ct}$ are properly chosen counter-terms. We will use this formula in the case of
potentials with a first order phase transition in order to compute the Free Energy difference between phases as a function
of temperature and determine the {\it critical temperature},  $T_c$, for those models. In evaluating this difference
the counter-terms will cancel that is why we do not need to have a detailed knowledge thereof. 

The way in which conformal symmetry is broken is determined by the choice of the scalar field potential which in our case is taken in a generic form \cite{Gubser:2008ny,Gursoy:2008za}
\begin{equation}
V(\phi)=-12\,(1+a\,\phi^2)^{1/4}\,\cosh(\gamma\,\phi)+b_2\,\phi^2 + b_4\,\phi^4 + b_6\,\phi^6.
\label{eq:V}
\end{equation}
The chosen potentials are summarized in table \ref{t.pot}.
Corresponding plots, representing temperature dependence of 
the entropy density, i.e., the equation of state (EoS), will be 
given together with the detailed discussion of each case 
in following sections.
Here we only make a few general remarks.
The parameters for the $V_{\rm QCD}$ potential
have been chosen to fit the lattice QCD (lQCD) data from ref. 
\cite{Borsanyi:2012cr}, and
the system is known to possess a crossover behaviour at zero
baryon charge density.
Parameters of potentials $V_{1\rm st}$ and $V_{2 \rm nd}$
were fitted so that the corresponding equations of 
state exhibit respectively the $1^{\rm st}$ , and the $2^{\rm nd}$ order 
phase transitions. In particular, for the $1^{\rm st}$ order case, in 
a certain temperature range 
we expect an instability 
(spinodal) region.

This concrete form of the last potential was already used explicitly in \cite{Gubser:2008yx} 
and is based on the considerations in \cite{Gursoy:2007cb} neglecting logarithmic running in the UV.
We will refer to it as the IHQCD potential 
\cite{Gursoy:2008za,Gursoy:2008bu}.
As it was mentioned in the introduction, and will be extended in 
section \ref{ihqcd}, it is designed to mimic some dynamical aspects of QCD.
However it is important to emphasize that the version used here is simplified as it does not
incorporate the UV logarithmic running.


\begin{table}
	\centering
	\begin{tabular}{c c c c c c c}
		\hline \hline
		\hspace{5pt}potential \hspace{5pt}& \hspace{5pt} $ a $ \hspace{5pt} & \hspace{5pt} $ \gamma $ \hspace{5pt} & \hspace{5pt} $ b_2 $ \hspace{5pt} & \hspace{5pt} $ b_4 $ \hspace{5pt} & \hspace{5pt} $ b_6 $ \hspace{5pt}  &  $ \Delta $  \\
		\hline
		\hspace{5pt}$V_{\rm QCD} $ \hspace{5pt}& 0 & \hspace{5pt}0.606 \hspace{5pt} & \hspace{5pt} 1.4 \hspace{5pt} & \hspace{5pt}-0.1 \hspace{5pt} & \hspace{5pt}0.0034\hspace{5pt} &\hspace{2pt}  3.55\hspace{5pt}  \\
		$V_{2\rm nd} $ &0  & $1/\sqrt{2} $ &  1.958 & 0  & 0 &  3.38  \\
		$V_{1\rm st} $ & 0 & $\sqrt{7/12}$  &  2.5  & 0  & 0 & 3.41  \\
		$V_{\rm IHQCD} $ & 1 & $\sqrt{2/3}$ & 6.25 & 0 & 0 & 3.58 \\
		
		\hline
	\end{tabular}
	\caption{Potentials chosen to study different equations of state
		exhibiting different phase structure and corresponding conformal dimension of the scalar field.
		\label{t.pot}}
\end{table}


The models determined by the potentials $V_{1\rm st}$ and $V_{\rm IHQCD}$ exhibit 
a first order phase transitions. In the former case the transition happens between
two different black hole solutions, while in the latter the transition happens
between a black hole and a horizon-less geometry. In both of those cases one can 
determine the transition by evaluating the FE difference according to formula
\eqref{eq:FE}, if one knows the counter terms
\footnote{Clarification
of this point can be found in the appendix \ref{AppB}.}. 
In this computation we follow an alternative method of ref. \cite{Gursoy:2008za} and integrate  the thermodynamic
relation, $d \mathcal{F}=-s\, d T$, with properly chosen boundary
condition. We can achieve this by first choosing some arbitrary
reference temperature $T_0$ and write
\begin{equation}
\mathcal{F}(T) = \mathcal{F}(T_0) - \int_{T_0}^T s(\tilde{T}) d\tilde{T}~,
\label{eq:Fint}
\end{equation}
where we assume to be in one particular class of solutions.
To evaluate the integration constant, $\mathcal{F}(T_0)$, we use the fact
that the Free Energy vanishes for the zero horizon area geometry. In general
the small horizon area limit of the black hole solutions corresponds
to the vacuum geometry with "good singularity" in the deep IR \cite{Gubser:2000nd}.
By using the relation of $T$ and $s$  and the horizon radius \eqref{entrodens}
we can evaluate the Free Energy with the data obtained with methods outlined in the
previous subsection. This amounts to 
a generic formula
\begin{equation}
\mathcal{F}(r_H) = -\frac{2\pi}{\kappa_5^2}\int_{r_H}^\infty \exp\left(3A(\lambda_H)\right)\frac{dT}{d \lambda_H} d\lambda_H~.
\label{eq:Frh}
\end{equation}
The details of the computations along with the corresponding plots and predictions
for $T_c$ will be given in the corresponding sections of the paper.


\section{Quasinormal modes}
\label{QNM}

In this section we formulate the problem of analyzing the linear 
perturbations around the equilibrium states in considered models.
The first subsection contains equations of motion and proper boundary
conditions that need to be imposed. The second subsection contains
a short summary of the results obtained with an emphasis on
generic aspects. The detailed case by case discussion
is a subject of the remaining part of the paper.


\subsection{ Equations of motion and boundary conditions}

The linear response of the system is analyzed by setting perturbations
with momentum in a given direction and computing poles of the resulting Green
functions. In this section we formulate the equations and corresponding boundary
conditions to present and discuss the results in the following part of the paper.

We consider perturbations of the background, obtained in the previous section, in the following form
\bea
&&g_{ab}(r, t, z) = g^{(0)}_{ab}(r) + h_{ab}(r)e^{-i\omega t + i k z}~,\\
&&\phi(r, t, z) = r + \psi(r)e^{-i\omega t + i k z}~. 
\eea
On the basis of \cite{Kovtun:2005ev,Benincasa:2005iv,Janik:2015waa} we consider infinitesimal
diffeomorphism transformations, $x^a\mapsto x^a +\xi^a$, 
of the form $\xi_a=\xi_a(r)e^{-i\omega t + i k z}$, 
which act on the perturbations in a standard way,
\be
g_{ab}\mapsto g_{ab} - \nabla_a\xi_b - \nabla_b\xi_a~, \hspace{35pt} \phi\mapsto \phi -\xi^a\nabla_a\phi~,
\ee
and look for linear
combinations of metric and scalar perturbations which are invariant under
those transformations.  There are four such modes, two of which are 
decoupled and two coupled.
Written explicitly, the coupled modes read
\begin{equation}
Z_1(r)=H_{aa}(r) \left(\frac{k^2 h'(r)}{2 A'(r)}+k^2 h(r)-\omega ^2\right)+k^2 h(r) H_{tt}(r)+\omega  (2 k H_{tz}(r)+\omega  H_{zz}(r))~,
\end{equation} 
and
\begin{equation}
Z_2(r) = \psi (r)-\frac{H_{aa}(r)}{2 A'(r)}~.
\end{equation}
In the above $h_{aa}(r)=h_{xx}(r)=h_{yy}(r)$ are transverse metric components and 
we have factorized the background from the metric perturbations in the following 
way:
$h_{tt}(r) = h(r)e^{2A(r)}H_{tt}(r),$ $h_{tz}(r) = e^{2A(r)}H_{tz}(r),$
$h_{aa}(r) = e^{2A(r)}H_{aa}(r),$  $h_{zz}(r) = e^{2A(r)}H_{zz}(r).$
Comparing with equation (3.12) of ref. \cite{Kovtun:2005ev} we can see
that $Z_1(r)$ mode corresponds to the sound mode, while the $Z_2(r)$ 
might be called a non-conformal mode, since it is intimately related to the 
scalar field.
The third  mode (which is decoupled) is the shear one and is expressed as 
\be
Z_3(r) = H_{xz}(r) + \frac{\omega}{k}H_{xt}(r)~,\label{shearMode}
\ee
and according to  the residual SO(2) symmetry in $xy$-plane (after turning on momentum along $z$-direction)  
is degenerated with the mode in which the index $x$ is replaced by the index $y$.
The dynamics of the fourth mode,
\be
Z_4(r)=H_{xy}(r)~,
\ee 
is governed by an equation
of motion which is similar to the external massless scalar 
equation, which was studied in with details in \cite{Janik:2015waa}.

The equations of motion for the modes $Z_1(r)$ and $Z_2(r)$ 
have the generic form
\bea
&&M_2(r) Z_1''(r)+M_1(r)Z_1'(r)+M_0(r)Z_1(r)+K_0(r)Z_2(r) = 0~,
\label{QNM1}\\
&&N_2(r) Z_2''(r)+N_1(r)Z_2'(r)+N_0(r)Z_2(r)+L_1(r)Z_1'(r)+L_0(r)Z_1(r) = 0~,
\label{QNM2}
\eea
and have to be solved numerically with proper boundary conditions.
The explicit form of the coefficient functions and comments about 
the numerics are given in the appendix \ref{app}.
As usual at the horizon we take the incoming condition,
which in our coordinates means the regular solution.

An analysis of the equations (\ref{QNM1}) and (\ref{QNM2})
near the conformal boundary leads to the asymptotic behavior
as $r\sim0$
\be\label{form}
Z_1(r)\sim A_1 + B_1\, r^{\frac{4}{4-\Delta}}~,
\hspace{40pt}
Z_2(r)\sim A_2\, r + B_2\, r^{\frac{\Delta}{4-\Delta}}~.
\ee
Transformation to the usual Fefferman-Graham coordinates close to the boundary, 
$r\mapsto\rho^{4-\Delta}$, reveals 
that $Z_1(\rho)$ has the asymptotic of 
metric components like the perturbations considered in \cite{Kovtun:2005ev}.
This perturbation corresponds to the sound mode of the theory.
On the other hand $Z_2(\rho)$ has the asymptotic of the 
background scalar field $\phi$ and is similar to the 
case studied in \cite{Benincasa:2005iv}.
The right boundary conditions for the QNM spectrum
are: $A_1=0$ and $A_2=0$.
The shear mode perturbation $Z_3(r)$ has the 
same asymptotic as $Z_1(r)$ and requires a 
standard Dirichlet boundary condition at $r=0$.


\subsection{General remarks and summary}
\label{remarks}

In all the cases the problem emerging from equations disused in the previous section
is a generalized eigenvalue
equation, which for a given $k$ results in a well defined
frequency $\omega(k)$. Note that all modes, for which 
$\textrm{Re }\omega(k)\neq0$, come in
pairs, namely
\be
\omega(k) =\pm\,|\textrm{Re }\omega(k)| + i\,\textrm{Im }\omega(k).
\label{pairRe}
\ee
As we will show in the next section in  some cases the modes are purely imaginary. 
But we want  to emphasize that in all of these cases (except the hydrodynamical shear mode) still we have pair modes with different values.
An important thing to note here is that
due to the coupled nature of the modes $Z_1(r)$ and $Z_2(r)$ there is another approximate degeneracy in the spectrum:
all modes, except for the hydrodynamical one, come in pairs. The reader is alerted not to confuse 
this structure with the one appearing in eq. (\ref{pairRe}).

For all the potentials we have made natural consistency checks.
For high temperatures (i.e., 
horizon radii closer to the asymptotic boundary)
in the sound and the shear channels we have an agreement with 
the pure gravity results dual to the CFT case  
\cite{Kovtun:2005ev}. 
The degeneracy related to the coupled nature of the modes is still present at high temperatures, where the 
system is expected to be conformal. The second most damped nonhydrodynamic
mode turns out to be the most damped nonhydrodynamic mode found in ref. \cite{Kovtun:2005ev}

The hydrodynamical QNM's are defined by the condition
$\lim_{k\rightarrow0}\omega_H(k)=0$, and are related
to transport coefficients in the following way
\be
\omega\approx-i\,\frac{\eta }{s\, T}\,k^2~,
\hspace{30pt}
\omega\approx\pm\, c_s\, k - i\,\Gamma_s\, k^2~,\label{dispersion-relations}
\ee
respectively in the shear and sound channels.
Those formulas are approximate in a sense that
in general higher order transport coefficients
should be considered \cite{Heller:2013fn}.
However, in a range of small momenta, second
order expansion is enough and we use it to read off
the lowest transport coefficients of the model.
The sound attenuation constant, $\Gamma_s$,
is related to shear $\eta$ and bulk $\zeta$
viscosities by
\be
\Gamma_s =
\frac{1}{2\,T}\,\left(\frac{4}{3}\,\frac{\eta}{s}+\frac{\zeta}{s}\right)~.
\ee
Those formulas were used to make the
second check of the results: compute the speed of sound 
$c_s$ and values of the shear viscosity  from the hydrodynamic 
modes and compare them respectively 
to the one obtained from the background calculations \eqref{Cs2} and predictions
known in the literature \cite{Kovtun:2004de,Janik:2006ft}. Both of them
are always satisfied, for example the classical result, $\eta/s=1/(4\pi)$ \cite{Kovtun:2004de},
is found in all cases considered in this
paper.

In classical gravity, the spectrum, apart from the hydro modes,
	contains of course also an infinite ladder of non-hydrodynamical modes. These are identified with the 
	poles of corresponding retarded Green's functions \cite{Kovtun:2005ev}, and as such correspond
	to physical excitations of the holographic field theory. In contrast to the hydrodynamic modes, we
	do not have a universal interpretation for them in gauge theory, however, this cannot stop us from
	treating them as physical excitations of the plasma system. Indeed, even if one is only interested
	in analyzing (high order) hydrodynamics, in \cite{Heller:2013fn}, one finds poles/cuts in the Borel plane
	which exactly correspond to the lowest \emph{non-hydrodynamic} QNM. This shows that these non-hydrodynamic 
	excitations have to be included for the self-consistency of the theory.
	
	Of course if one is close to equilibrium, the higher QNM will be more damped and may be neglected
	in practice. However in some cases the lowest QNM become comparable to the hydrodynamic ones and
	as such provide an applicability limit for an effective hydrodynamic description. These phenomena
	will be at the focus of the present paper. Indeed we find that they become very important in
	the vicinity of a phase transition.
	
	Finally, to demystify somewhat these higher quasinormal modes, one can give a well known simple physical setup
	when only these modes are relevant. Suppose that one considers a spatially uniform plasma system
	and starts with an anisotropic momentum distribution for the gluons. Then the initial energy-momentum tensor
	is spatially constant but anisotropic. If we let the system evolve, the system will thermalize (with the
	energy-momentum tensor becoming eventually isotropic). However this (homogeneous) isotropization will not
	excite any hydrodynamic modes as the symmetry of the problem forbids any flow. Thus the relevant
	excitations will be different. At strong coupling they correspond exactly to the higher quasinormal
	modes.

In the analysis below we measure the momenta and the frequencies in the units of temperature
by setting 
\be
q=\frac{k}{2\pi T}, 
\hspace{40pt}
\varpi= \frac{\omega}{2 \pi T} ~.
\ee
There are a few novel predictions which we make from the QNM
frequencies. First is to estimate the momentum, or equivalently the length,
scale where the hydrodynamic description of the system breaks.
For a CFT case this was evaluated to be $q=1.3$ where in the shear 
channel first non-hydrodynamic QNM dominated the system dynamics \cite{Landsteiner:2012gn}.
In the same time this effect did not appear in the CFT sound channel. 
The new feature we find is that we see this crossing\footnote{In this paper by crossing between the modes we mean crossing in the imaginary part of the hydrodynamic and the most damped non-hydrodynamic modes.} not only
in the shear channel but also in the sound channel. This shows that 
the influence of a non-trivial phase structure of the background 
affects the applicability of hydrodynamics in a qualitative way. 
Other aspect is that 
the hydrodynamic description is valid in large enough length scale (the smaller critical momentum)
which means the applicability of hydrodynamics 
near the phase transition is more restricted than in the high 
temperature case. 

In table \ref{t.qc} we summarize the critical momenta  in two channels and hydrodynamic parameters for different potentials. All quantities are evaluated at corresponding critical temperatures.
In the following subsections we will show the QNM's mostly for the sound channel which present characteristic structure for each potential. Since the shear channel in all cases has the same form (with different critical momentum) we restrict ourselves to show only one related plot for the $V_{\rm QCD}$ potential.


\begin{table}[h!]
	\centering
	\begin{tabular}{|c| c | c | c | c |}
		\hline \hline
		\hspace{10pt}	potential \hspace{10pt} & \hspace{10pt}  sound channel $q_c$ \hspace{10pt}   & \hspace{10pt} shear channel $q_c$ \hspace{10pt} &\hspace{17pt} $c_s^2$ \hspace{17pt} &\hspace{10pt}  $\zeta/s$ \hspace{17pt}\\
		\hline
		$V_{\rm QCD} $ & $0.8$  &  $1.1$  & $0.124$ & $0.041$  \\
		$V_{2\rm nd} $ & $0.55$ &  $0.9$ & $0.0$ &  $0.061$   \\
		$V_{1\rm st} $ & $0.8$  &  $1.15$ & $0.0$ & $0.060$ \\
		$V_{\rm IHQCD} $ & $0.14$  &  $1.25$ & $0.0$ & $0.512$ \\	
		\hline
	\end{tabular}
	\caption{The momenta for which the crossing phenomena in different channels and the corresponding
		values of the speed of sound and bulk viscosity read of from the hydrodynamic mode. Values given at corresponding critical temperatures
		($T_{\rm m}$ for $V_{1{\rm st}}$ and $V_{\rm IHQCD}$).	
		\label{t.qc}}
\end{table}


The second observation is the bubble formation in the spinodal 
region in the case of the $1^{\rm st}$ order phase transition \cite{Chomaz:2003dz}.
This happens when $c_s^2<0$ which means that hydrodynamical mode
is purely imaginary.
For small momenta, $\omega_H=\pm i |c_s| k - i \Gamma_sk^2$, the mode with the plus sign is in the unstable region,
i.e., ${\rm Im} ~\omega_H>0$. For larger momenta the other term starts 
to dominate, so that there is $k_{\rm max}=|c_s|/\Gamma_s$ for which the hydro mode
becomes stable again. The scale of the bubble is the momentum for which
positive imaginary part of the hydro mode attains the maximal value.
Imaginary part of the unstable hydro mode is called the growth rate \cite{Chomaz:2003dz}.

Third observation is that 
the hydrodynamical mode of the sound channel in $1^{\rm st}$ order case near the critical temperature $T_c$, and in the IHQCD case also the first non-hydrodynamical modes, become purely imaginary
for a range of momenta. Interpretation of this fact is that the corresponding wavelengths
cannot propagate at a linearized level and correspondingly there is a diffusion-like
mechanism for those modes. 

It is important to note that generically the ultra-locality \cite{Janik:2015waa}
of the non-hydrodynamic mode is still present in the critical region of the phase diagram.
The only exception observed is the IHQCD potential, where the modes exhibit a non
trivial behaviour. Most of the interesting dynamics and effects observed are due to the
different behaviour of the hydrodynamical modes and how they cross the most damped non-hydrodynamic modes. This includes the instability and the 
bubble formation in the case of the $1^{\rm st}$ order  phase transition.


\section{The crossover case}
\label{secVQCD}

The results for the QNM with a QCD-like equations of 
state are summarized below. Parameters of this potential
have been chosen to fit the temperature dependence of the 
speed of sound obtained in
lattice QCD computations with dynamical quarks at zero
baryon chemical potential \cite{Borsanyi:2012cr}.


\begin{figure}
	\begin{center}
		\includegraphics[height=.23\textheight]{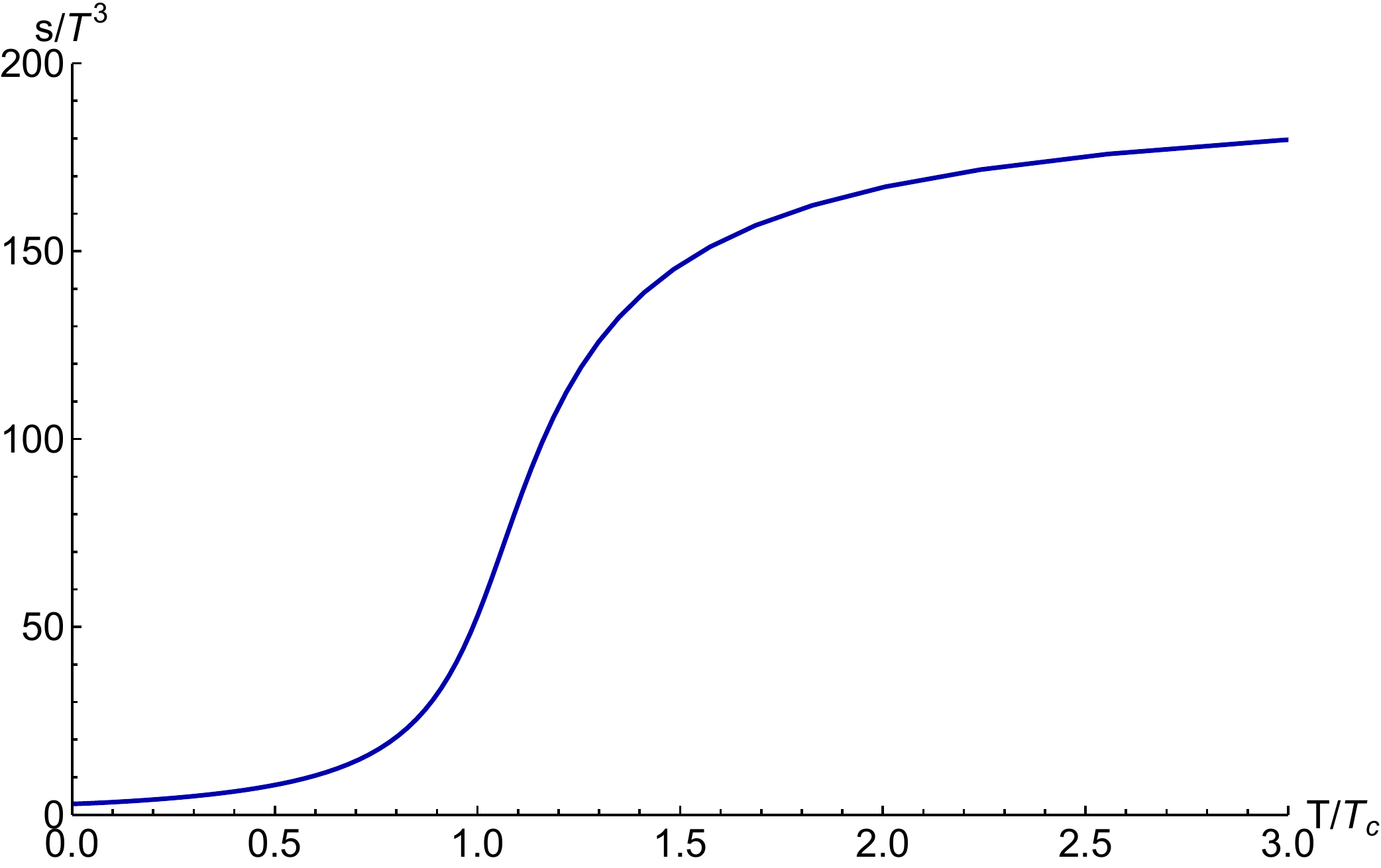}
		\includegraphics[height=.23\textheight]{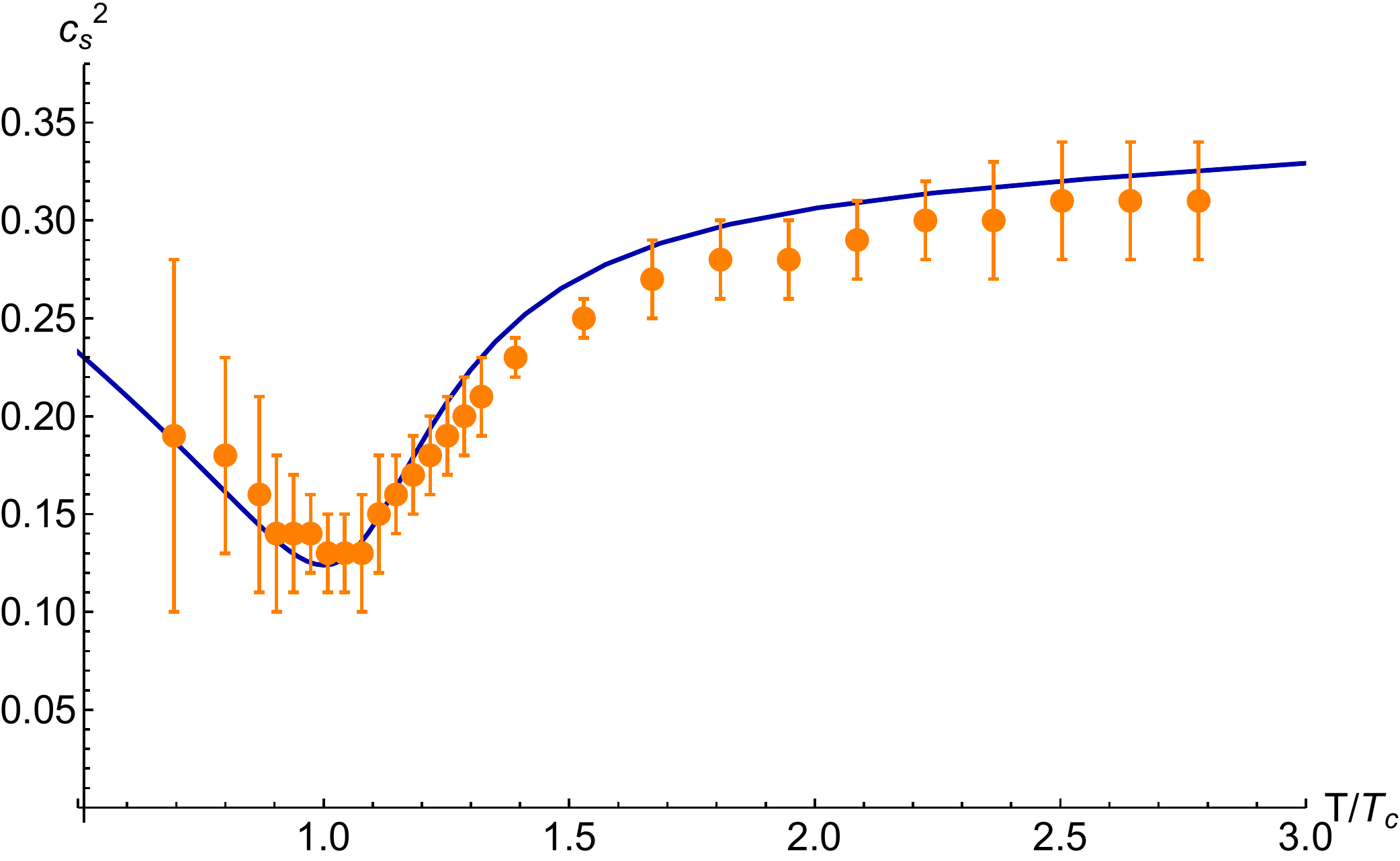}
		\caption{Left Panel: Entropy density for $V_{\rm QCD}$ potential with $\kappa_5=1$. Right panel: Speed of sound squared as a function of temperature. Dots are the lQCD equation of state \cite{Borsanyi:2012cr}.}
		\label{EoS(VQCD)}
		\end{center}
\end{figure}


In our computations from the hydrodynamic mode we estimate the value 
of the bulk viscosity, which is in agreement with ref. \cite{Gubser:2008yx} (cf. tab. \ref{t.qc}).
It is important to note, that despite the fact that the EoS of QCD
are correctly reproduced in the model transport coefficients are lower than the lattice predictions
\cite{Meyer:2007dy,Karsch:2007jc}. 
For example only the qualitative temperature dependence of bulk viscosity is correct, namely that it rapidly raises 
near the $T_c$ \cite{Gubser:2008yx}.\footnote{We define the pseudo-critical temperature as the lowest value for the speed of sound \eqref{Cs2}. Corresponding lQCD definition refers to peaks of chiral and Polyakov loop susceptibilities  \cite{Bazavov:2009zn,Borsanyi:2010bp}.}

In this analysis we take another step, and study the temperature and momentum behaviour
not only of the hydrodynamic mode but also of the first and second of the 
infinite tower of higher modes. In particular this allows us to estimate
the applicability of the hydrodynamic approximation in the critical region
of temperatures where we find crossing of the modes in sound channel.

Firstly, before we move to the new results, using the example of the $V_{\rm QCD}$ potential, let us 
discuss the high temperature quasinormal modes.
The results computed for $T=3T_c$ are shown in
fig. \ref{VQCD(hT-QNM)}. The speed of sound, shear and bulk viscosities
read of from the lowest QNM are very close to results 
expected for a conformal system, i.e., 
$
\eta/s\simeq1/(4\pi), \hspace{5pt} c_s^2\simeq0.321, \hspace{5pt} \zeta/s\simeq0.003.
$
Modes computed for this 
temperature in the sound
and the shear channels are in  agreement with the conformal results 
of ref. \cite{Kovtun:2005ev}.
As we mentioned in previous section, since $Z_1(r)$ and $Z_2(r)$ modes are coupled 
the nonhydrodynamic QNM's are in pair in all range of temperatures,
and the second most damped nonhydrodynamic mode turns out to be the most damped one found in ref. \cite{Kovtun:2005ev}.


\begin{figure}[t!]
	\begin{center}
		\includegraphics[height=.22\textheight]{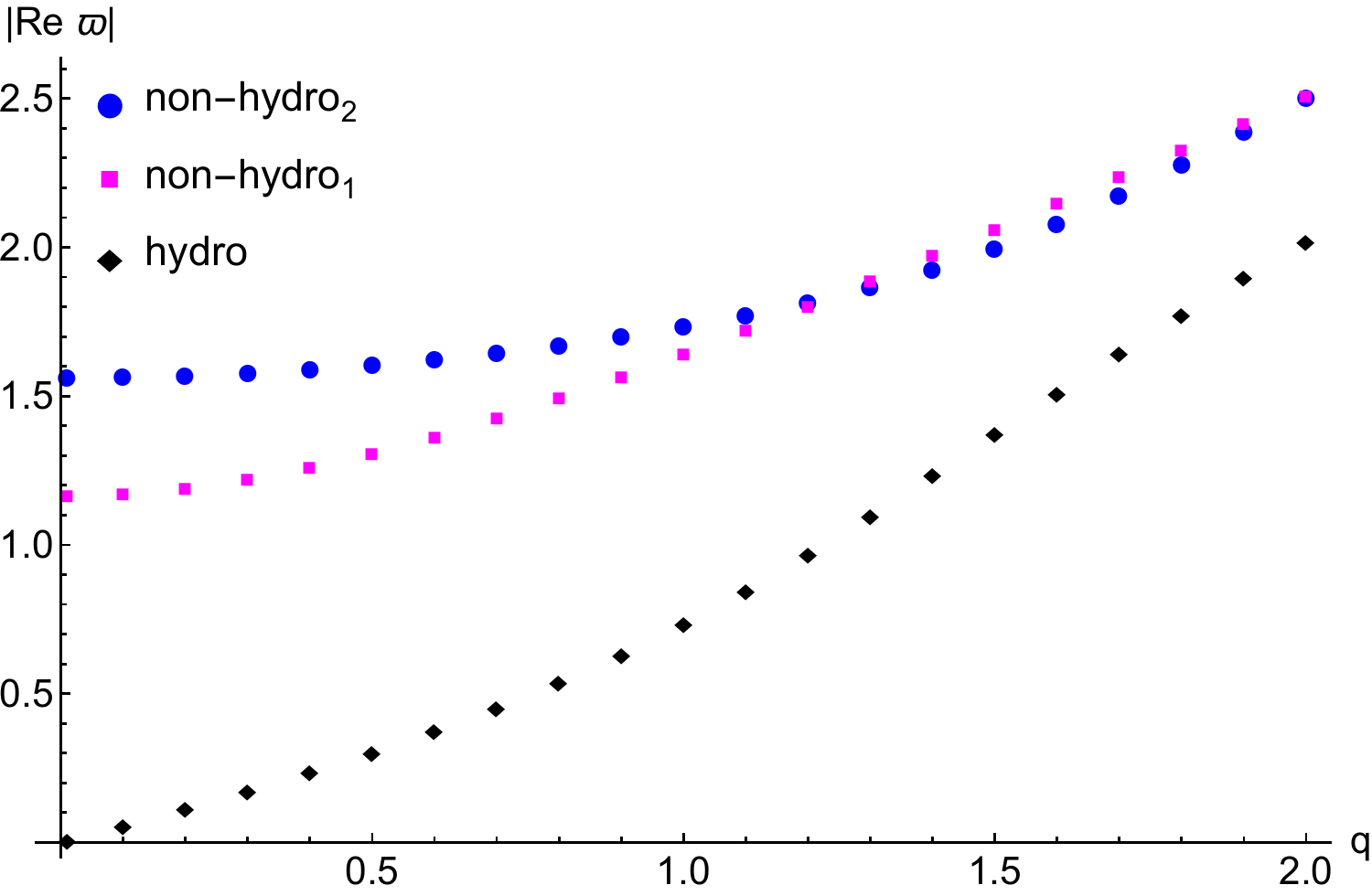}
		\includegraphics[height=.22\textheight]{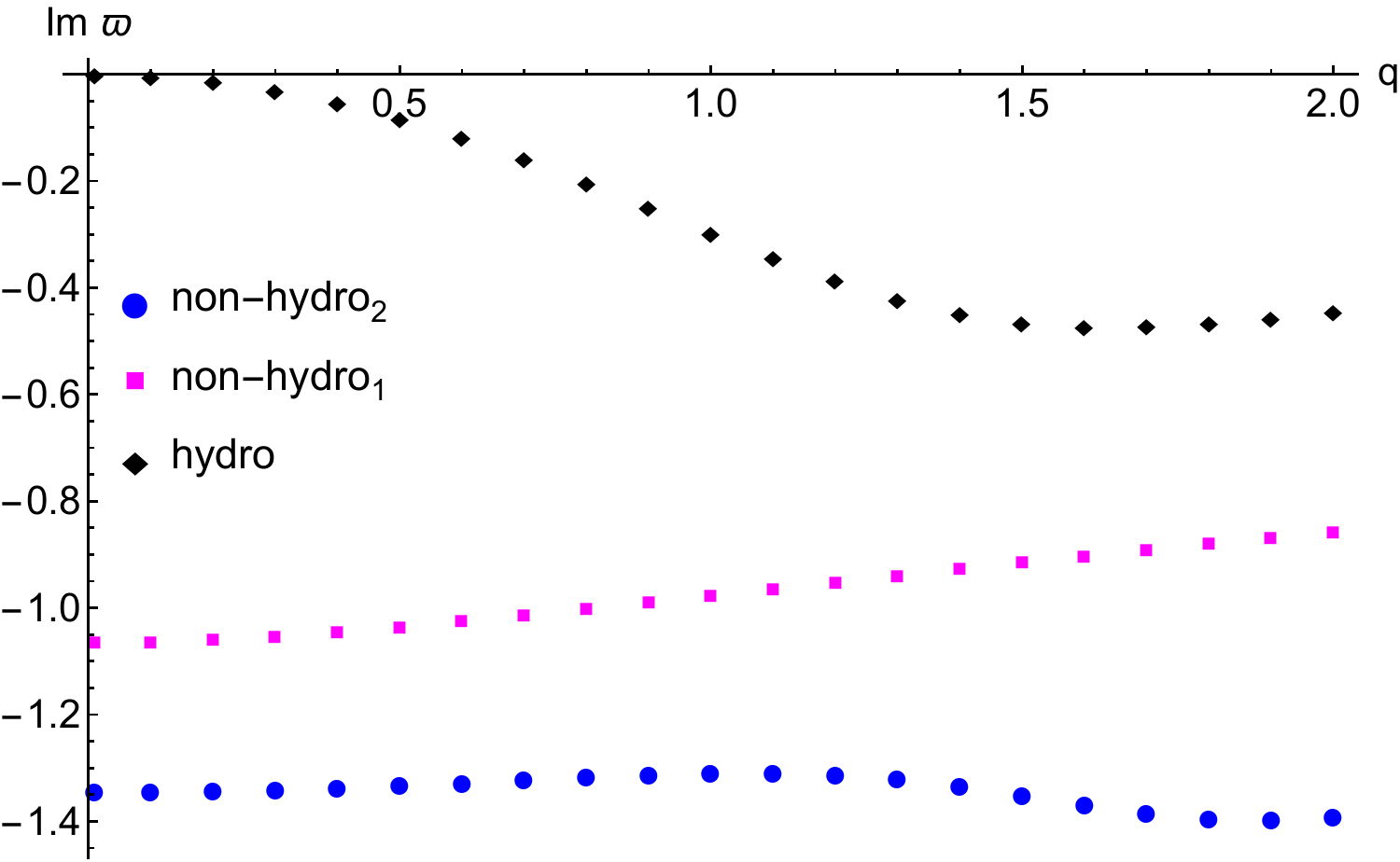}
		\caption{Sound channel quasinormal modes for the potential $V_{\rm QCD}$ at $T=3T_c$.
			Real part (left panel) and imaginary part (right panel).
			}
		\label{VQCD(hT-QNM)}
	\end{center}
\end{figure}



\begin{figure}[h!]
	\begin{center}
		\includegraphics[height=.22\textheight]{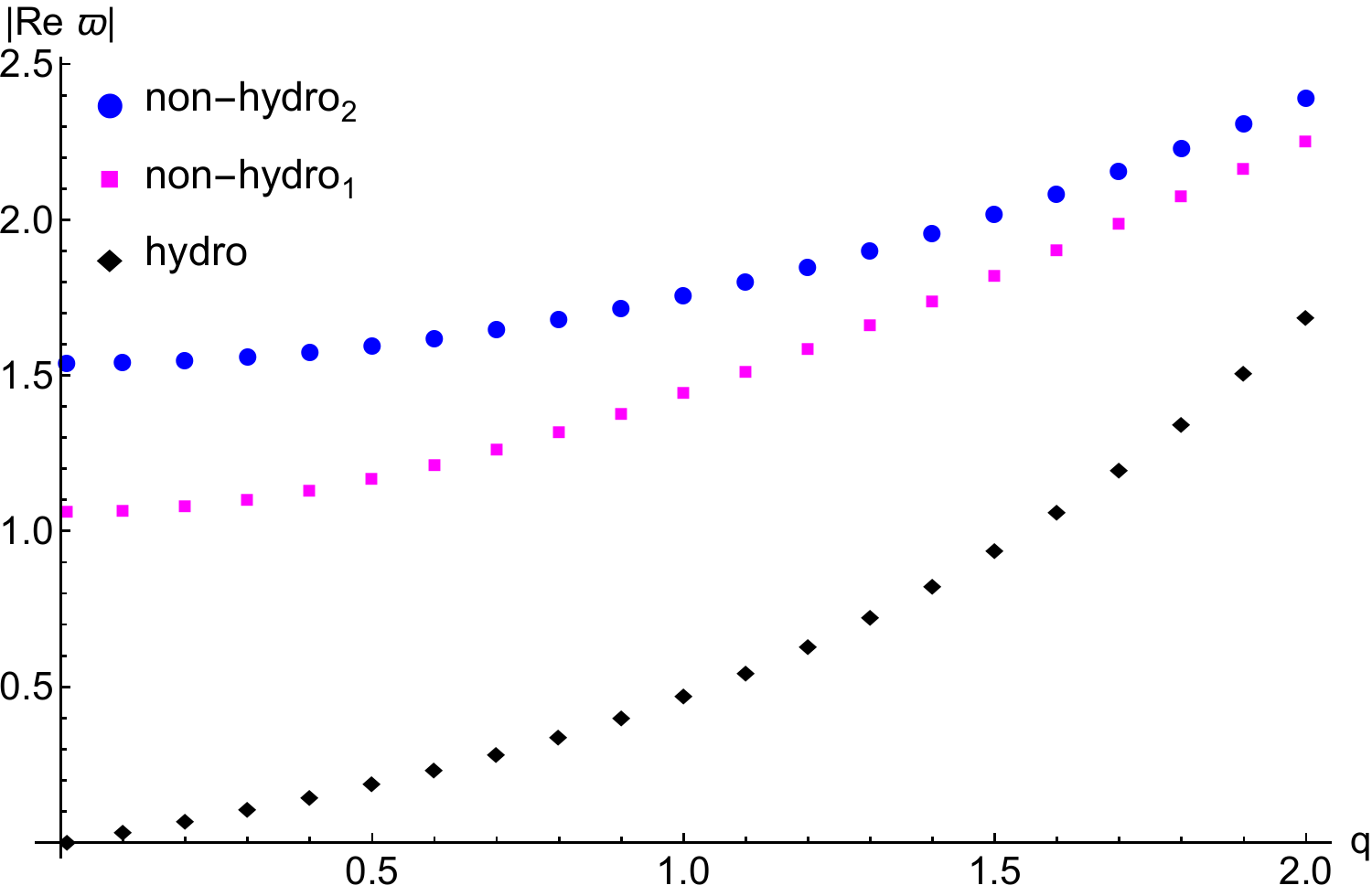}
		\includegraphics[height=.22\textheight]{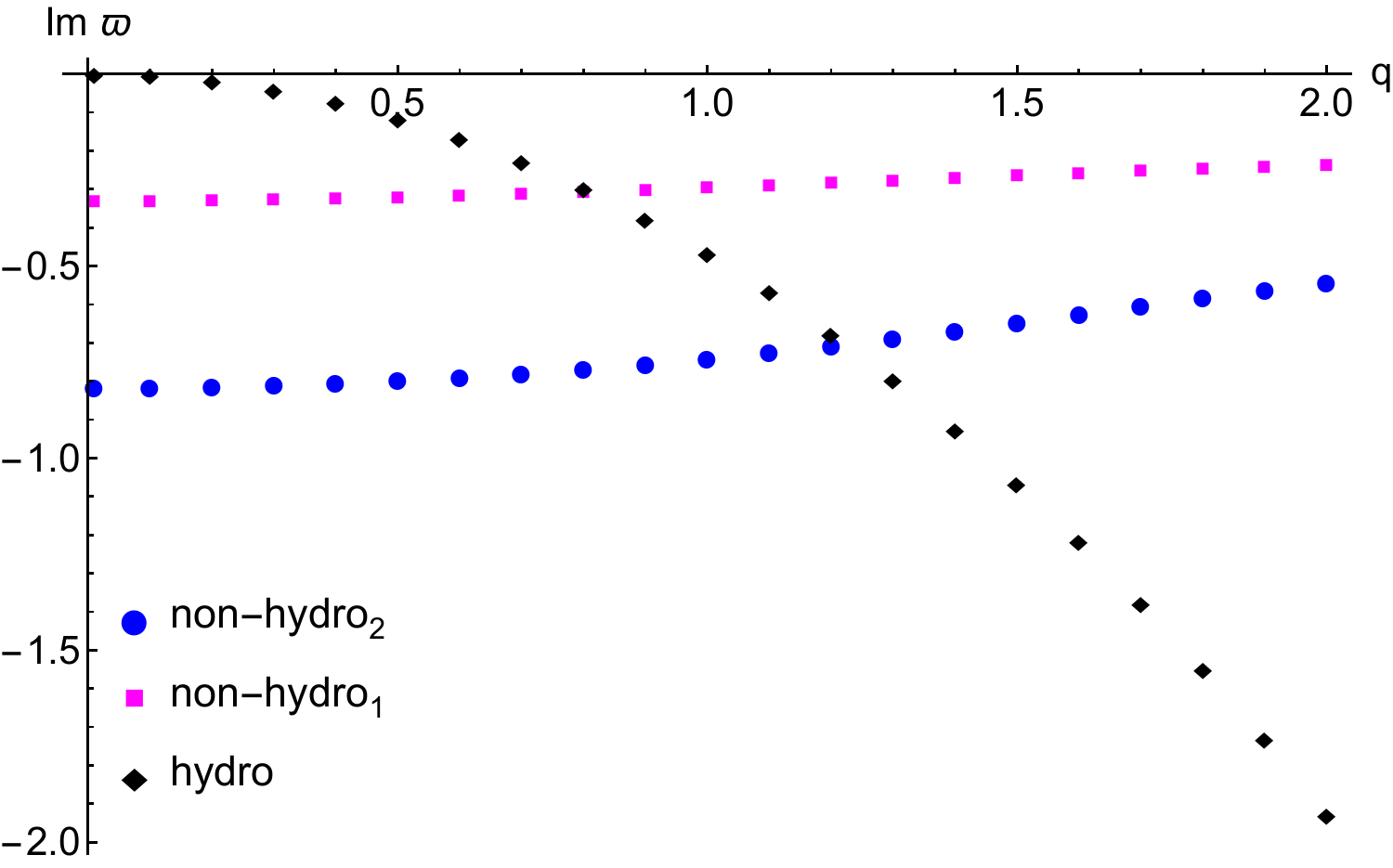}
		\caption{The real (left) and imaginary (right) parts of the quasinormal modes in the sound 
		channel for the potential $V_{\rm QCD}$ at  $T=T_c$. 
		}
		\label{VQCD(QNM)}
	\end{center}
\end{figure}

\begin{figure}[h!]
	\begin{center}
		\includegraphics[height=.22\textheight]{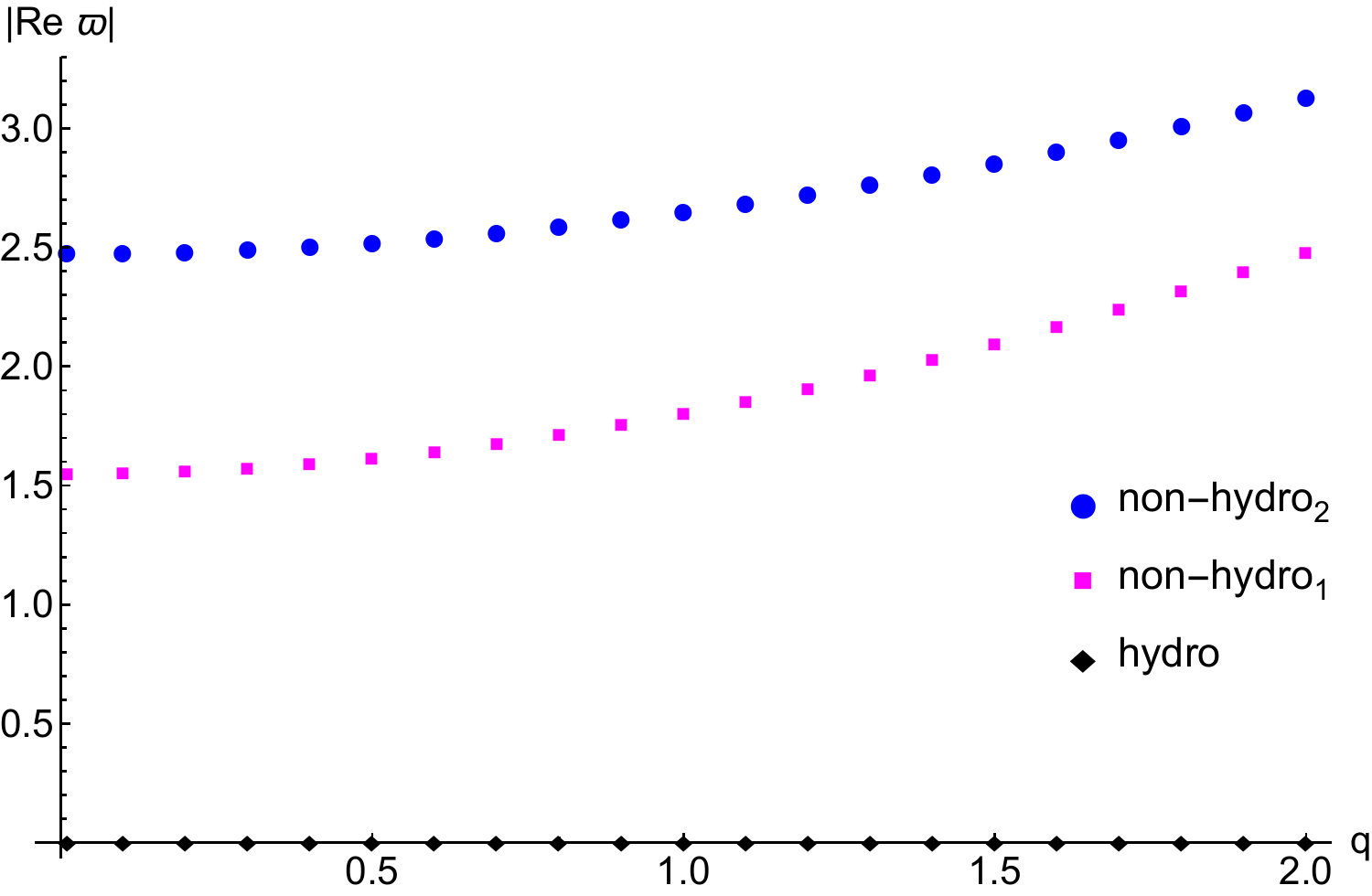}
		\includegraphics[height=.22\textheight]{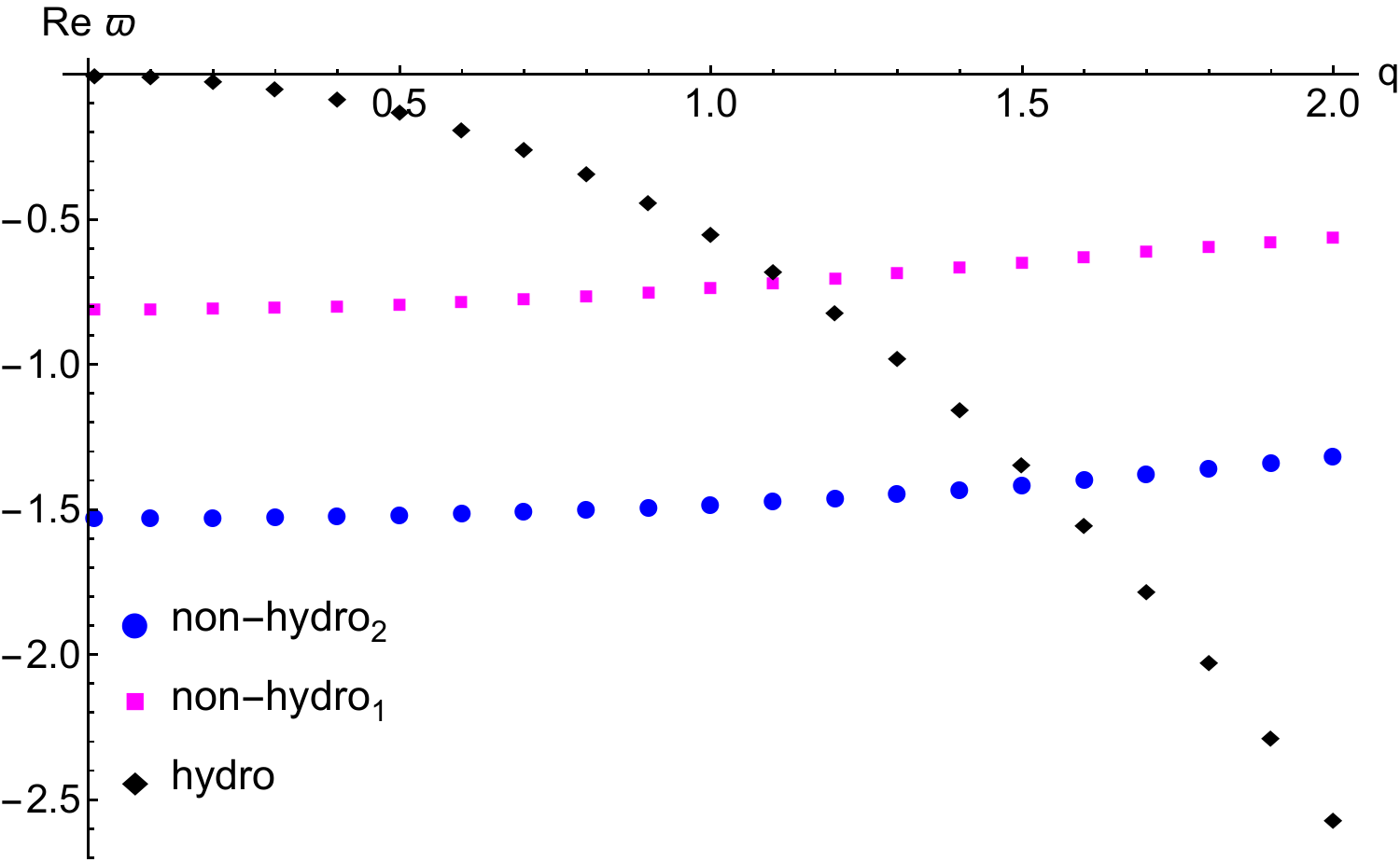}
		\caption{The real (left) and imaginary (right) parts of the quasinormal modes in the shear 
		channel for the potential $V_{\rm QCD}$ at  $T=T_c$.
		}
		\label{VQCD(shear)}
	\end{center}
\end{figure}

Now let us turn our attention to the opposite case of lower temperatures.
The results computed for the pseudo-critical temperature, $T=T_c$,
are shown in fig. \ref{VQCD(QNM)}. The most important difference with respect
to high-$T$ case a change in large momentum dependence of the imaginary part 
of the hydrodynamic mode.
Instead of approaching some constant value the imaginary
part of the mode flows to minus infinity as momentum increases. 
This implies a novel effect in the sound channel: crossing between the 
hydrodynamic and non-hydrodynamic mode appears.  At the pseudocritical
temperature this happens for critical momentum $q_c\simeq0.9$. 
While in the conformal case this was present only in the shear channel for  $q_c\simeq1.3$ \cite{Landsteiner:2012gn}, as shown in figure \ref{VQCD(shear)} for the crossover potential $q_c\simeq1.15$ in the same channel.
In contrast, nonhydrodynamic modes are not much affected obeying ultra locality
property \cite{Janik:2015waa}. 


In view of possible relations to QCD we could expect
only qualitative predictions from our computations.
However, lattice QCD computations could, in principle, verify 
the ultra-locality property and the generic
crossing of the modes. The main obstruction in 
this case would be the necessity of real time 
formulation of the problem, which is not yet 
available on the lattice.


\section{The second order phase transition case}
\label{2nd}

\begin{figure}[h!]
	\begin{center}
		\includegraphics[height=.23\textheight]{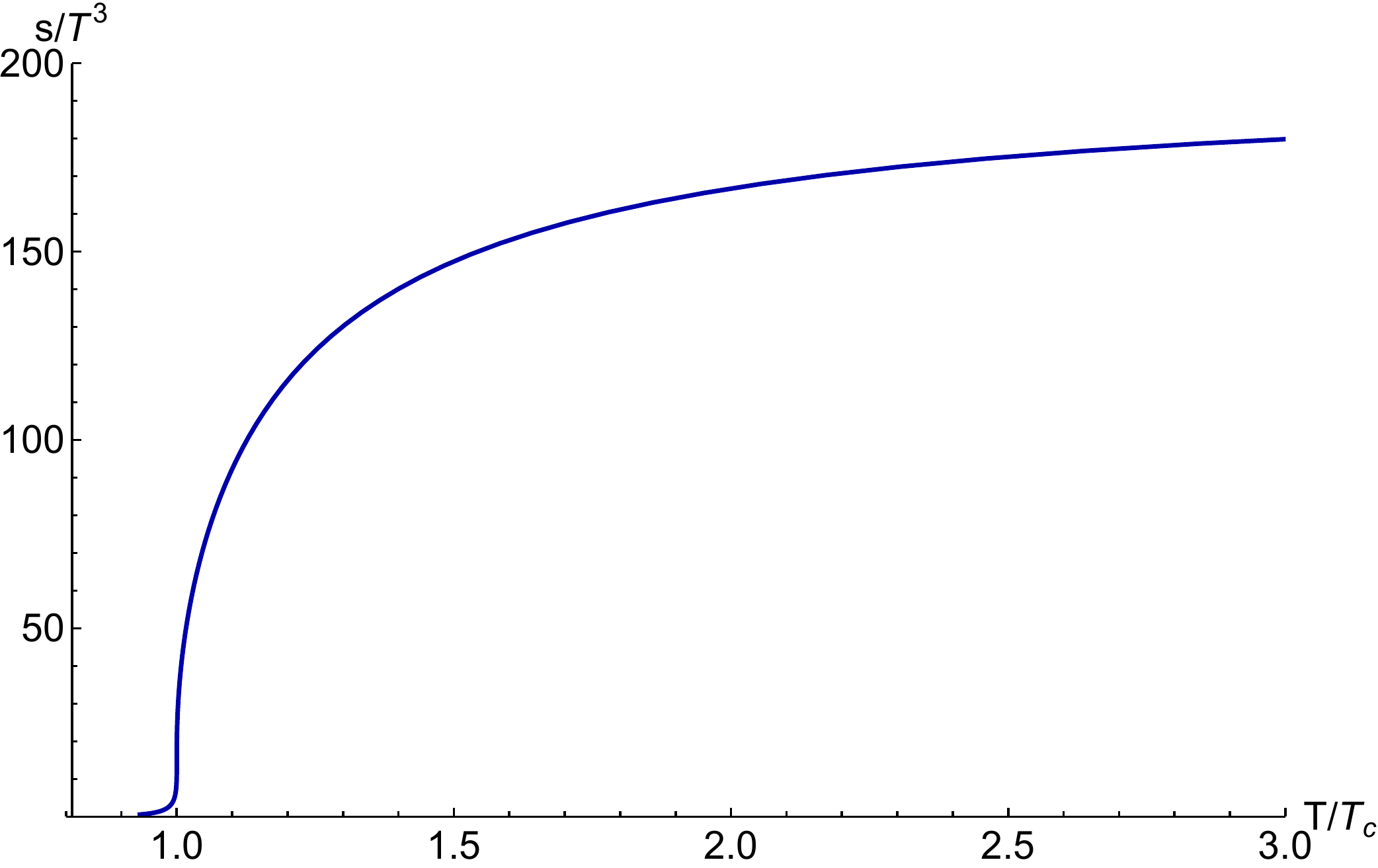}
		\includegraphics[height=.23\textheight]{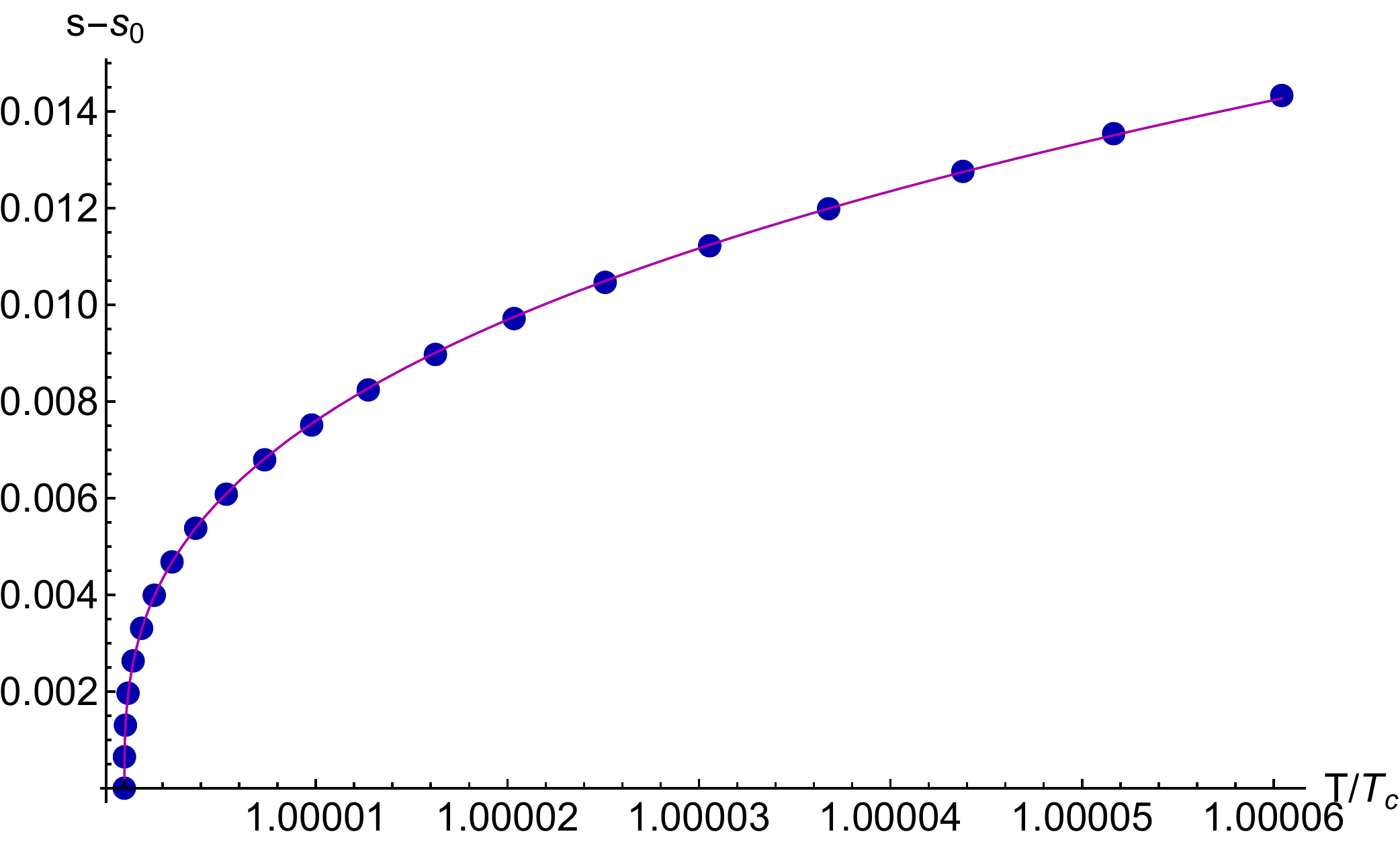}
		\caption{Left panel: equation of state for $V_{2\rm nd}$.
			 Right panel: Equation of state for $V_{2\rm nd}$ near the $T_c$ (blue points).
			Magenta line is the fit \eqref{eq:Fit} with $\alpha\simeq0.65$.
			In both plots we set $\kappa_5=1$.
		}
		\label{V2nd(alpha)}
	\end{center}
\end{figure}

In this section we present results for the case of a system
with $2^{\rm nd}$  phase transition EoS, which can be achieved 
by a suitable choice of parameters. We do not fit to any particular 
system considered in the literature - we only require a particular
shape of the entropy as a function of temperature 
(cf. left panel of fig. \ref{V2nd(alpha)}) which leads to vanishing speed of sound at the critical temperature $T=T_c$
\cite{Gubser:2008ny}. Near the $T_c$ entropy of the system takes the form
\begin{equation}
s(T)\sim s_0 + s_1 t^{1-\alpha}~,
\label{eq:Fit}
\end{equation}
where $t=(T-T_c)/T_c$, and $\alpha\simeq0.65$ is 
the specific heat critical
exponent (cf. right panel of fig. \ref{V2nd(alpha)}). This value is very close to 
$\alpha=2/3$ from ref. \cite{Gubser:2008ny}.


\begin{figure}[h!]
	\begin{center}
	\includegraphics[height=.22\textheight]{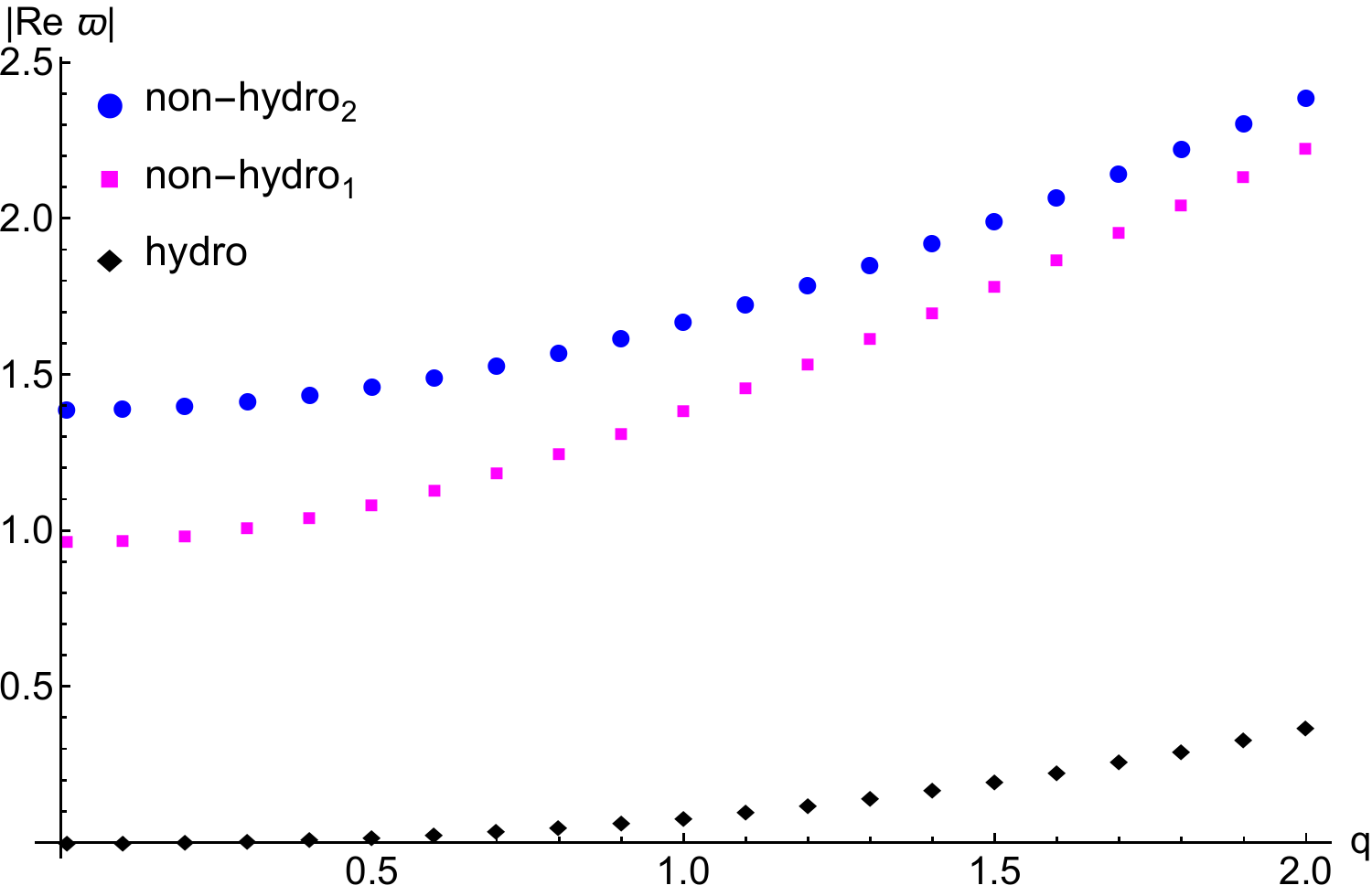}
	\includegraphics[height=.22\textheight]{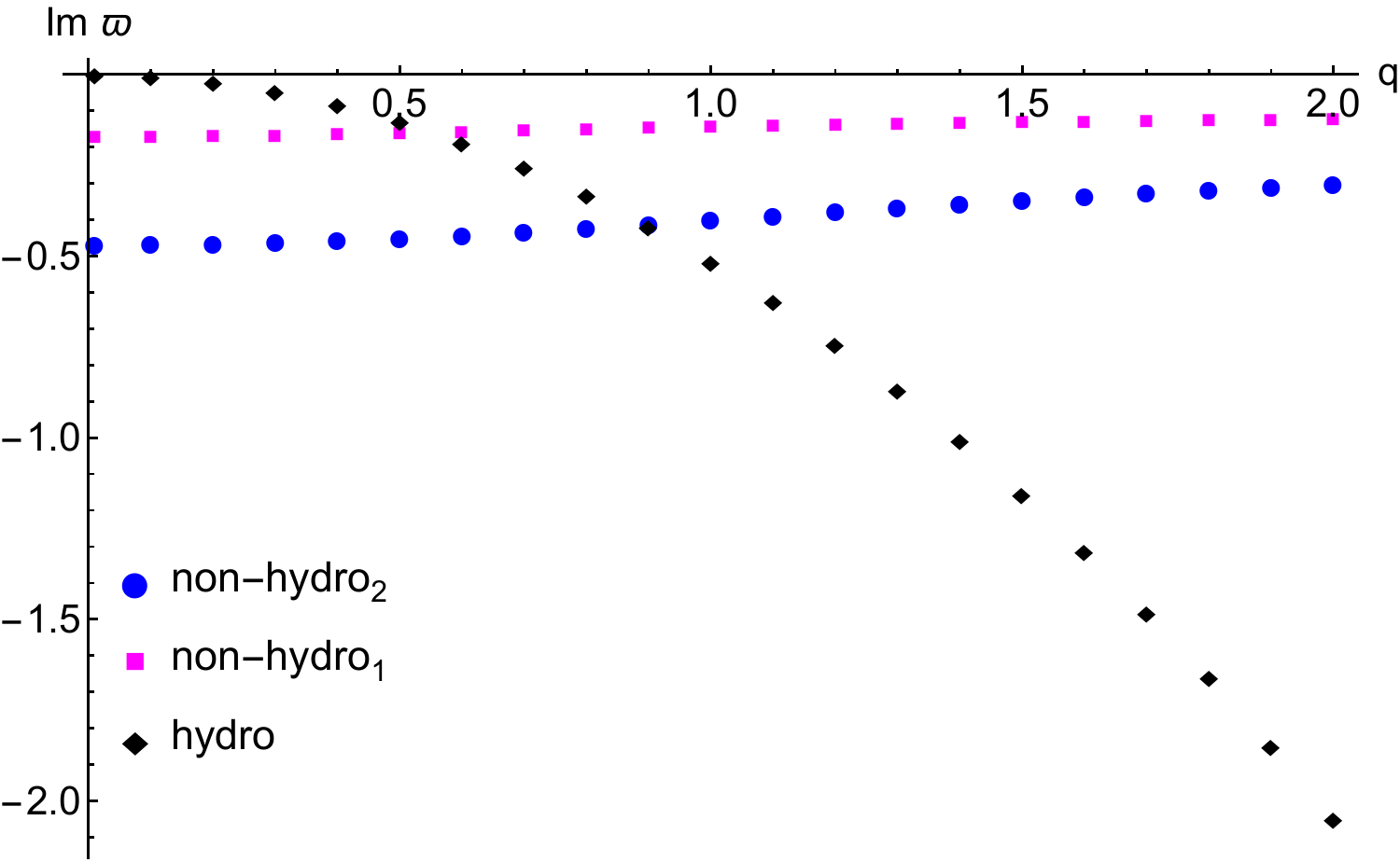}
	\caption{Quasinormal modes for the potential $V_{2\rm nd}$ at $T_c$.
		Real part (left panel) and imaginary part (right panel).
		}
	\label{V2nd(QNM)}
	\end{center}
\end{figure}

The results for QNM at critical temperature 
are displayed in fig. \ref{V2nd(QNM)}. 
Since there is no new phenomena in the shear channel, only the sound mode is shown. Generic temperature dependence of QNM 
frequencies is very similar to the crossover case.
The main difference compared to the crossover potential (fig. \ref{VQCD(QNM)}) is that at $T_c$ the hydrodynamic description
of the system breaks down already at smaller momenta scales.


We would like to mention that in high temperature regime we recovered the CFT results in both channels
 with the pair structure explained in the previous subsection in the sound channel due to coupling of the modes.


\section{The first order phase transition case}
\label{1st}

In this section we discuss the most  fascinating case of a system
which exhibits a $1^{\rm st}$ order phase transition. There are two possible scenarios
for such a transition: one is similar to Hawking-Page case where there is a 
transition from a black hole to the vacuum geometry without a horizon \cite{Hawking:1982dh}.
The second one, mentioned in ref. \cite{Gursoy:2008za}, is a transition from one black hole solution to another.
In this section we consider the latter case while the former appears in the studies of IHQCD models
(cf. sec. \ref{ihqcd}). The onset of the appearance of a nonpropagating sound mode in the deeply 
overcooled phase has been observed earlier in a related model \cite{Gursoy:2013zxa}.


\begin{figure}
	\begin{center}
		\includegraphics[height=.23\textheight]{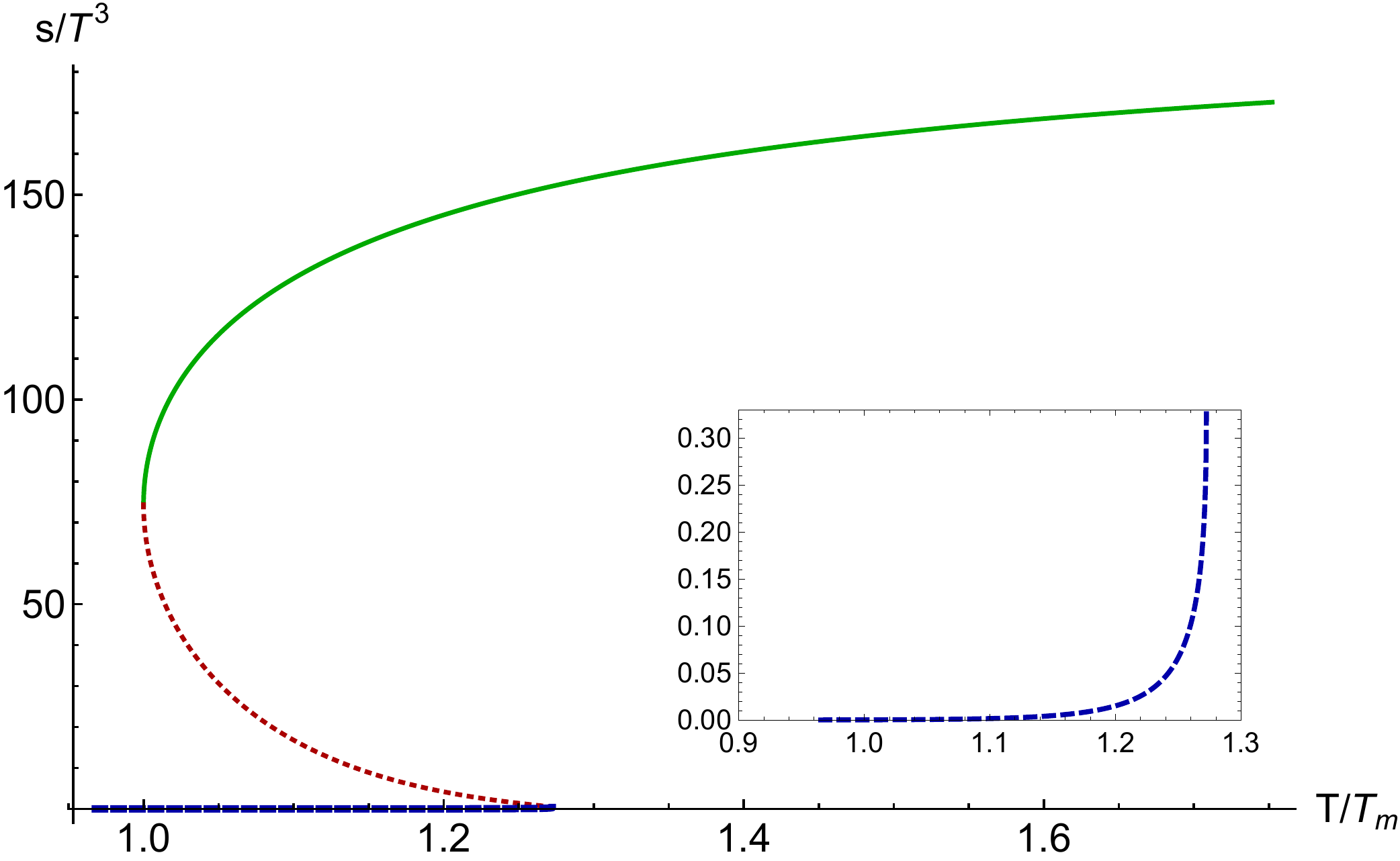}
		\includegraphics[height=.23\textheight]{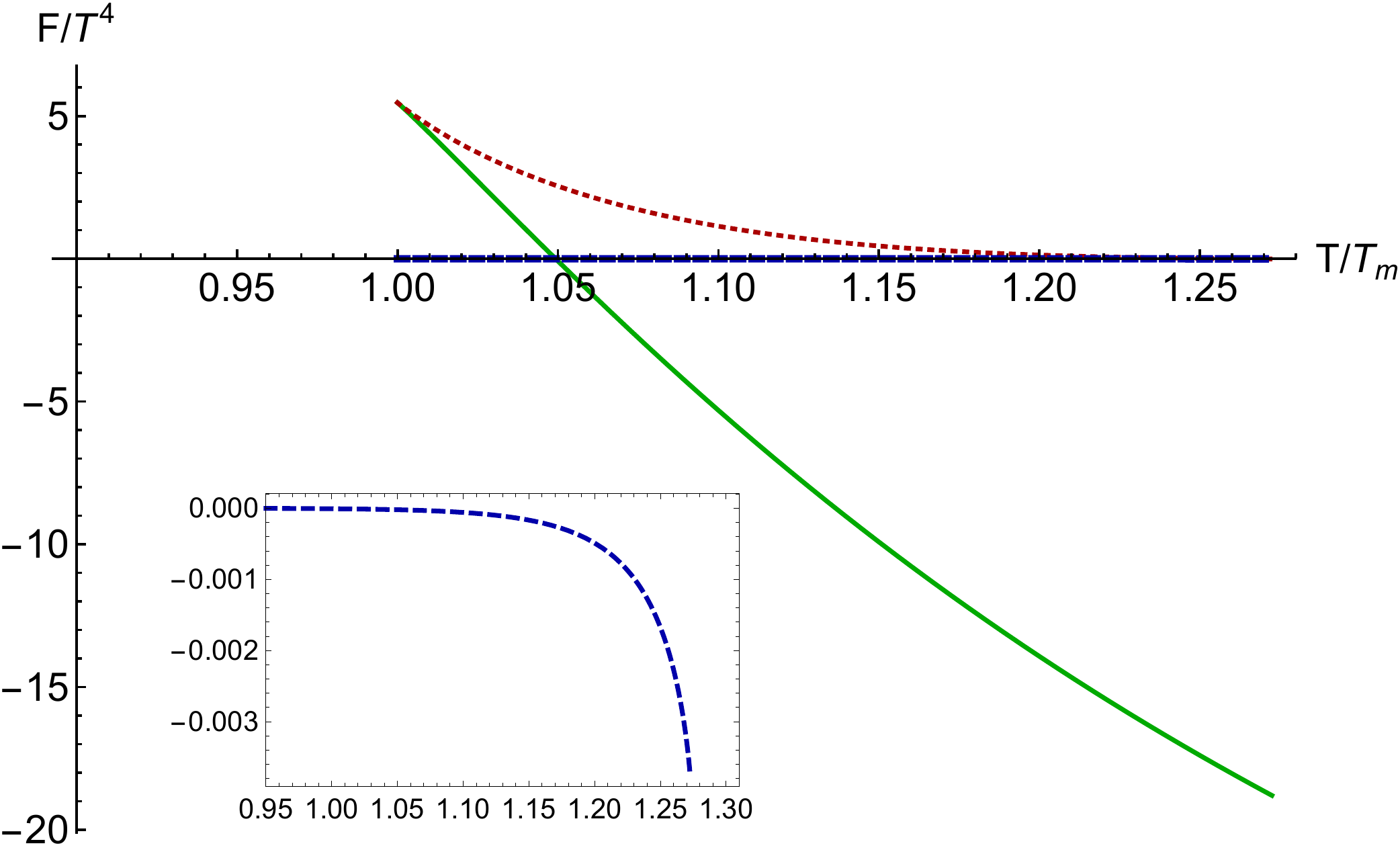}
		\caption{Left panel: Entropy density for $V_{1\rm st}$ potential. Green line is the stable region, 
			while red dashed line displays an instability. Right panel: Free Energy difference between two black hole solutions as a function of
			temperature. Estimated critical temperature is $T_c\simeq 1.05T_m$. 
			In both plots we set $\kappa_5=1$.}
		\label{EoS(V1st)}
	\end{center}
\end{figure}

\begin{figure}[t!]
	\begin{center}
	\includegraphics[height=.22\textheight]{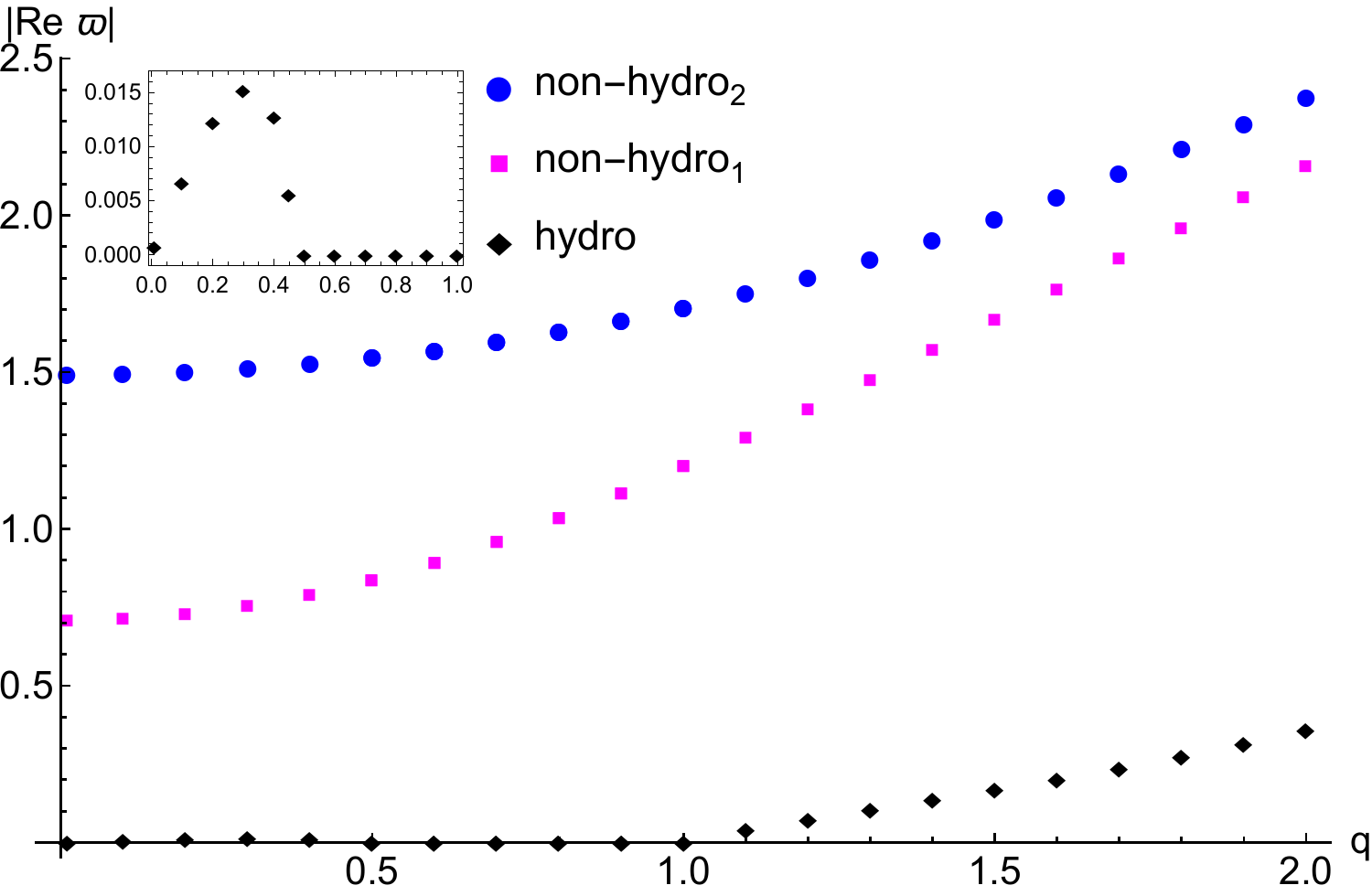}
	\includegraphics[height=.22\textheight]{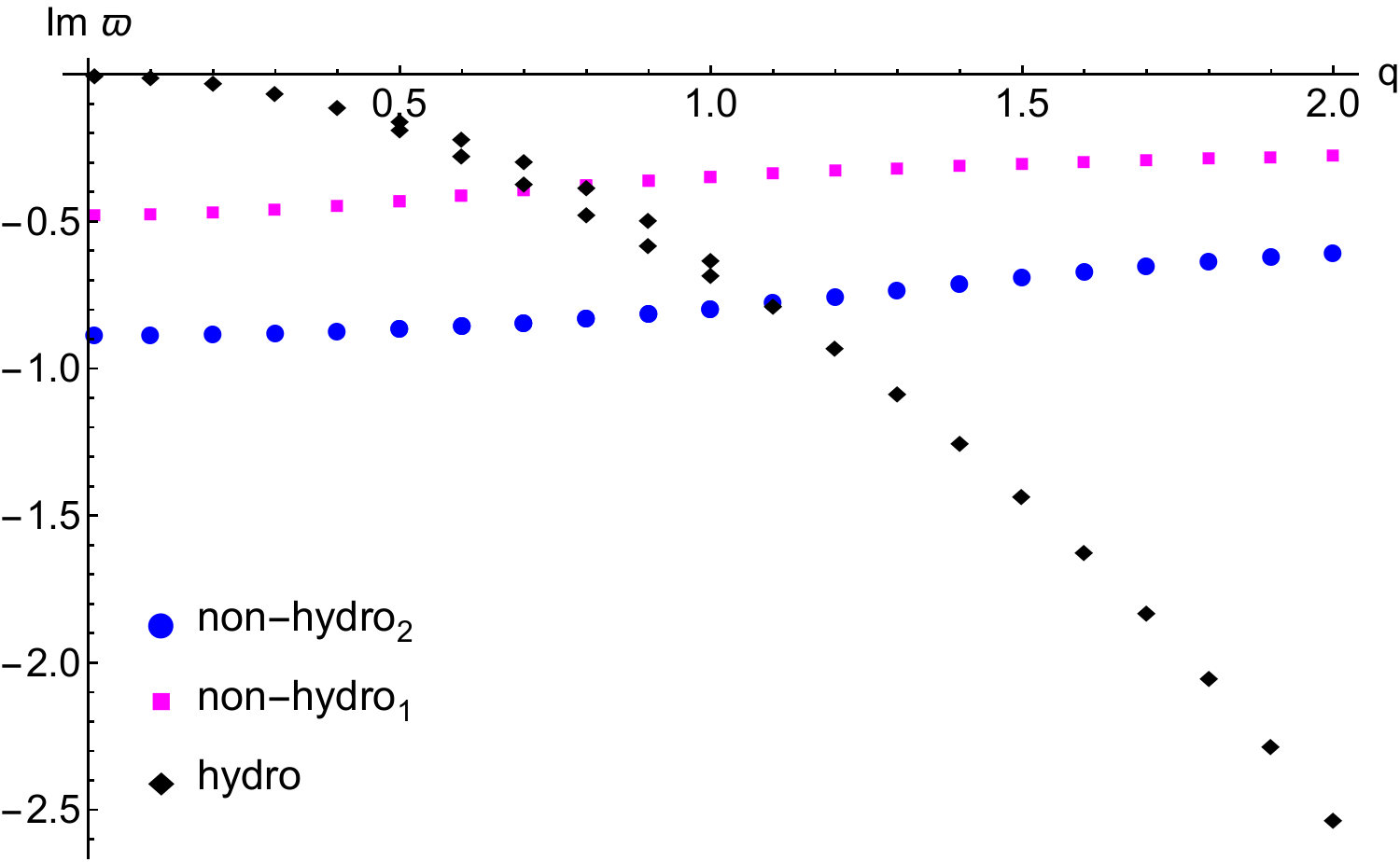}
	\caption{Quasinormal modes for the potential $V_{1\rm st}$ at $T=1.00004T_m$. 
		Real part (left panel) and imaginary part (right panel).}
	\label{V1st(QNM-2)}
	\end{center}
\end{figure}

In the $V_{1\rm st}$ potential case there exist three characteristic temperatures. The first one is the minimal temperature $T_m$, below which no unstable solution exists. 
The onset of instability is seen at temperatures $T\gtrsim T_m$ (in the branch where $c_s^2(T)<0$), and generically we expect the ${1\rm st}$ 
order phase transition to appear at a critical temperature  $T_c\geq T_m$,
which is determined by
the temperature dependence of the Free Energy.
To evaluate this one can either use direct on-shell actions or one can use the method outlined in section \ref{Background}.
The latter uses the standard thermodynamic relation $d\mathcal{F}=-s\,dT$, where the integration constant can be fixed by the choice of the reference geometry with vanishing horizon area, which in this case corresponds to $T=0$ solution.
Temperature dependence of the FE for this case is shown in the right panel of fig. \ref{EoS(V1st)} and we determined $T_c\simeq 1.05T_m$.
The other characteristic temperature is estimated to be $T_{\rm ch}\simeq 1.0001 T_m$,
which is based on the observation, that for a range of momenta the hydrodynamic modes become purely imaginary and do not propagate in the plasma (cf. fig \ref{V1st(QNM-2)}).  
This effect appears for temperatures $T_m\leq T\leq T_{\rm ch}$, in the \emph{stable} 
region of the EoS (green line in left panel of fig. \ref{EoS(V1st)}). Let us note that
in this model $T_m<T_{\rm ch}<T_c$.


\begin{figure}[b!]
	\begin{center}
	\includegraphics[height=.22\textheight]{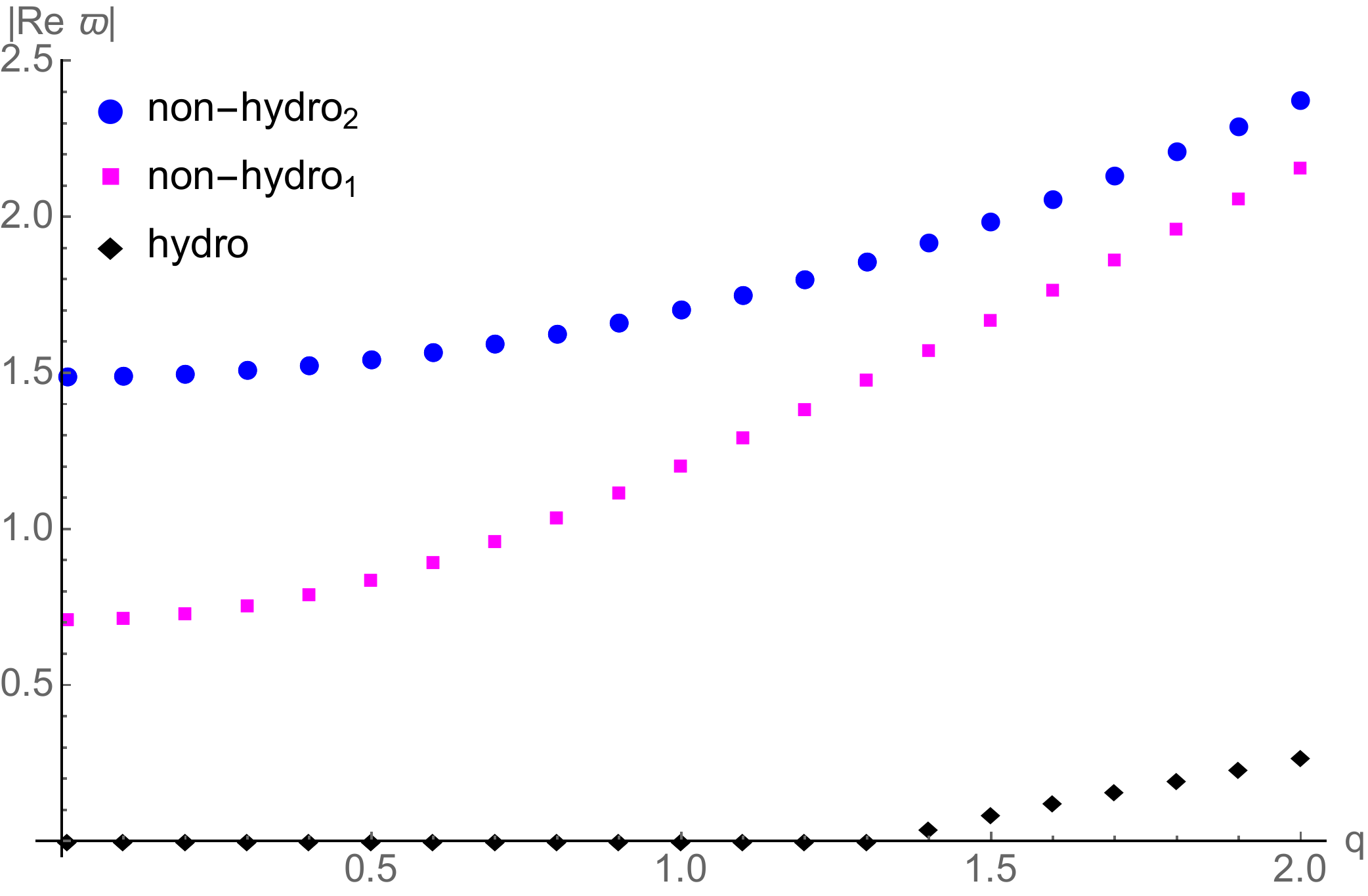}
	\includegraphics[height=.22\textheight]{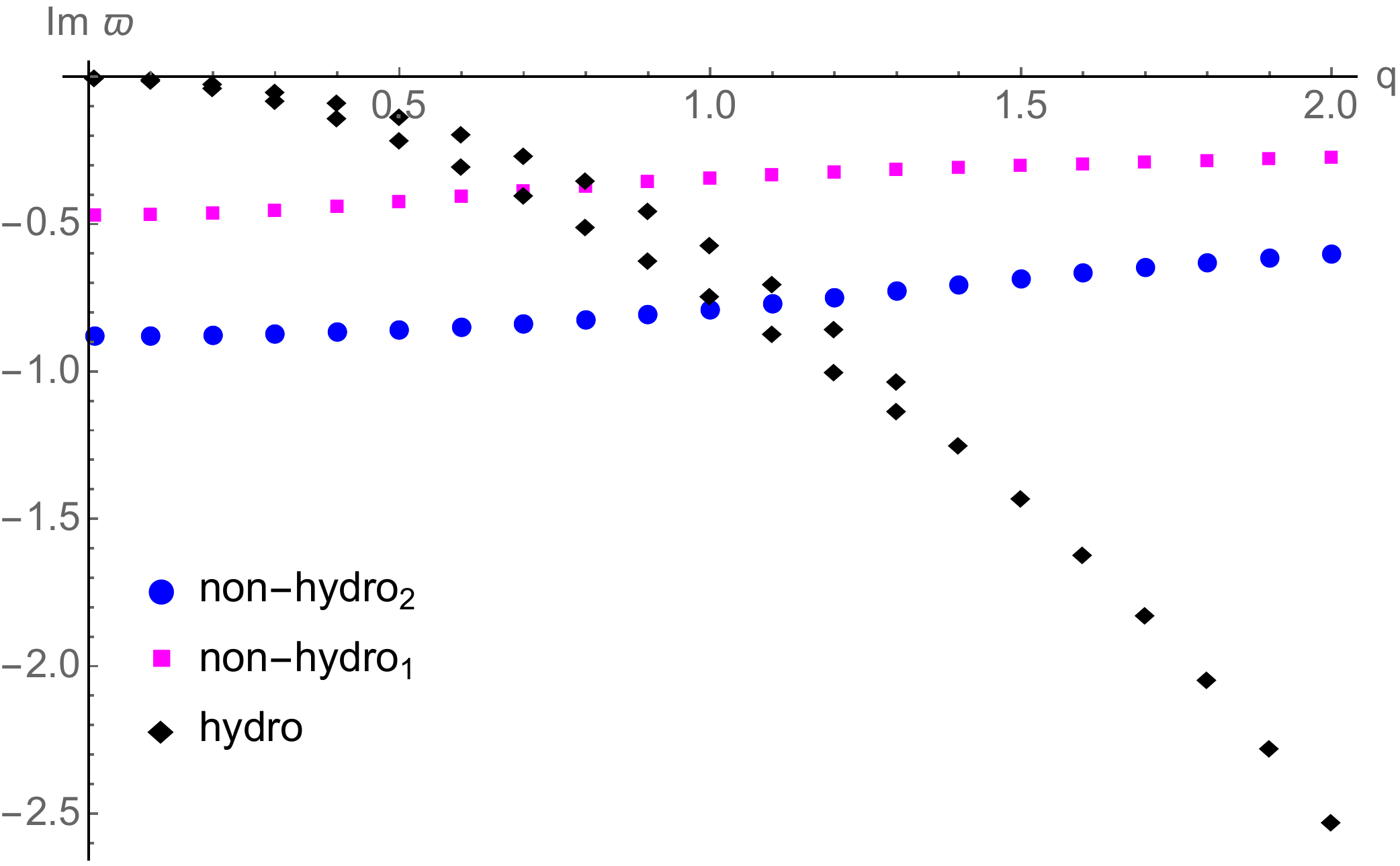}
	\caption{Quasinormal modes for the potential $V_{1\rm st}$ at $T=T_m$. 
		Real part (left panel) and imaginary part (right panel).}
	\label{V1st(Tc)}
	\end{center}
\end{figure}

Now we take a look at QNM structure at the minimal temperature $T_m$, in which the green 
line and red-dashed line meet in fig. \ref{EoS(V1st)} and the speed of sound vanishes. 
There is no new structure in the shear channel and we only plot the sound channel QNM's in figure \ref{V1st(Tc)}. 
One may see a new pattern at this point compared to the crossover and the $2^{\rm nd}$ order  phase transition cases, i.e., 
the hydrodynamic modes are purely imaginary (diffusive-like) for $q\leq1$.


The most engrossing physics is discovered 
in the \emph{spinodal} region (red-dashed line fig. \ref{EoS(V1st)}) where the equation of state 
suggests thermodynamical instability, i.e., $c_s^2<0$
(cf. fig. \ref{EoS(V1st)}). It was 
already anticipated in literature \cite{Gubser:2000ec,Buchel:2005nt} that in 
this range of temperatures a corresponding dynamical instability should appear in the lowest QNM mode.

\begin{figure}[t!]
	\begin{center}
\includegraphics[height=.22\textheight]{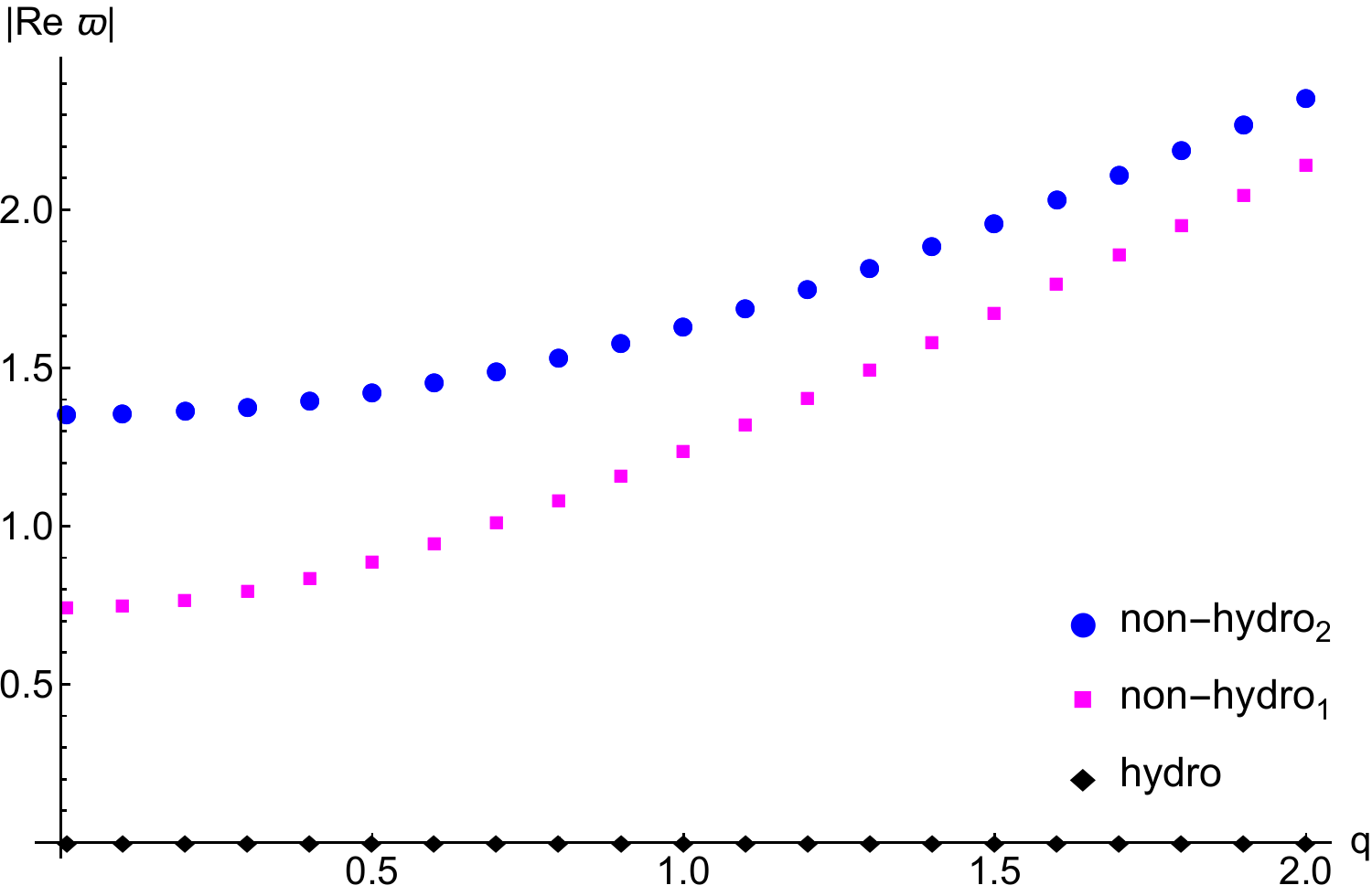}
\includegraphics[height=.22\textheight]{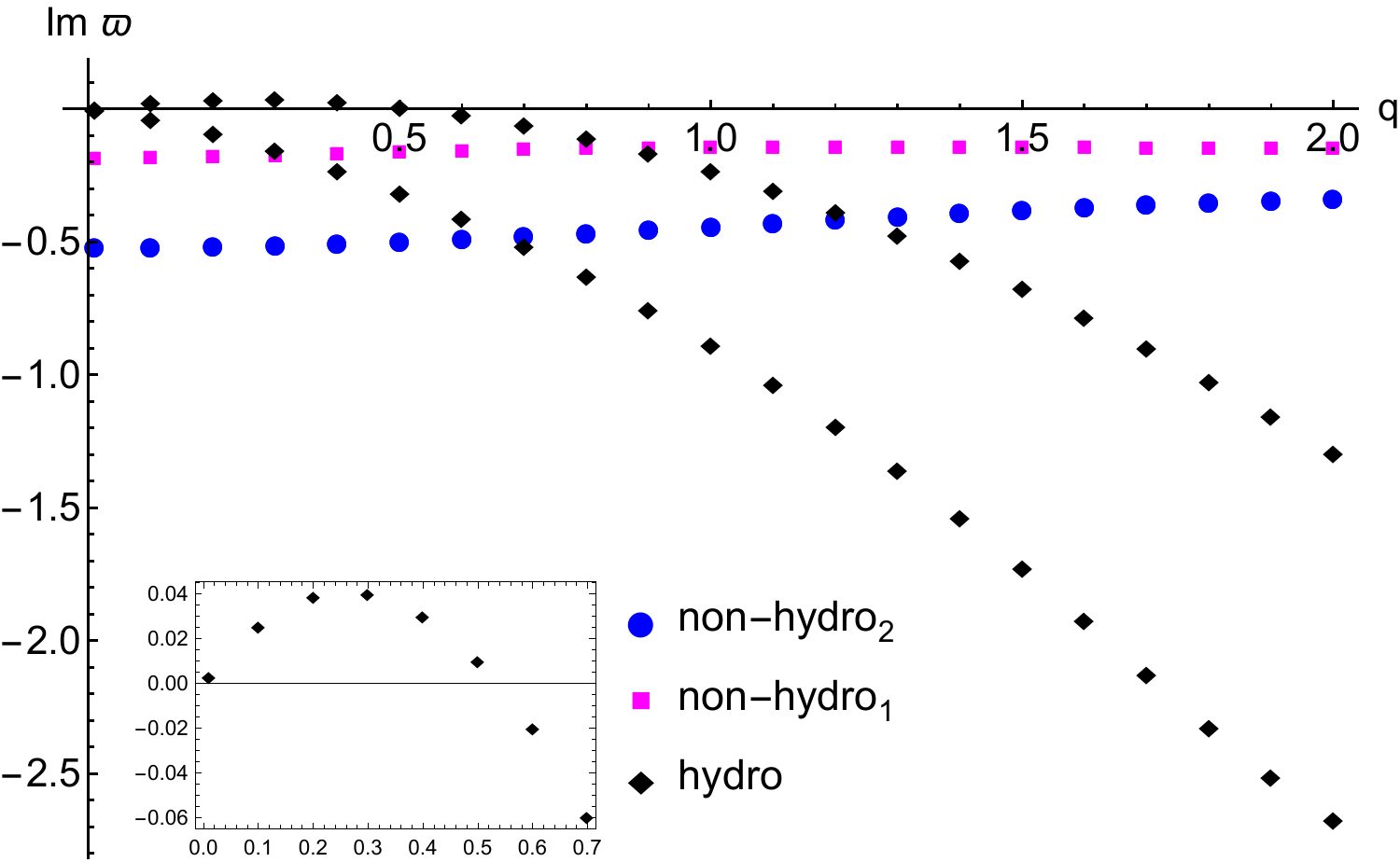}
\caption{Sound channel quasinormal modes for the potential $V_{1 \rm st}$ at  $T\simeq 1.06 T_m$.
An instability of the spinodal region is shown. The speed of sound at that temperature is
$c_s^2\simeq-0.1$.
}
\label{V1st(QNM)}
\end{center}
\end{figure}

We study the instability phenomenon in detail by observing the bubble formation in the spinodal 
region. It is generically expected in the case of the $1^{\rm st}$ order phase transition \cite{Chomaz:2003dz}
and a similar effect was observed in the gravity context by Gregory and Laflamme \cite{Gregory:1993vy}.
The formation happens when $c_s^2<0$, which means that hydrodynamic mode
is purely imaginary $\omega_H=\pm i |c_s| k - i \Gamma_sk^2$.
For small enough $k$ the mode with the plus sign is in the unstable region,
i.e., ${\rm Im} ~\omega_H>0$. For larger momenta the other term starts 
to dominate, so that there is $k_{\rm max}=|c_s|/\Gamma_s$ for which the hydrodynamic mode
becomes again stable. The scale of the bubble is the momentum for which
positive imaginary part of the hydrodynamic mode attains the maximal value.
Imaginary part of the unstable hydrodynamic mode is called the growth rate \cite{Chomaz:2003dz}.
It is intriguing  to note, that for the $V_{1\rm st}$ potential the hydro
mode is purely imaginary up to momenta $q\approx5$, i.e., in all investigated range.
An interesting observation is that all higher modes remain stable in this case.
Plot illustrating these words is presented in fig. \ref{V1st(QNM)}.


Not only in the case when there is an instability region in the  EoS, but also for the temperature close to the $T_m$ in the stable region and at $T_m$
the hydrodynamic modes become purely
imaginary. These cases are shown in figures \ref{V1st(QNM)}, \ref{V1st(QNM-2)}, \ref{V1st(Tc)} respectively.
When a hydrodynamic mode, $\omega_H(k)$, is purely imaginary, one can express it as
\begin{equation}
\omega_H(k)=\pm i O(k) - i E(k)~,
\label{OddEven}
\end{equation}
with $O(-k)=-O(k)$ and $E(-k)=E(k)$.
Then there are two separated branches of the hydrodynamical modes,
as seen in figures \ref{V1st(QNM-2)}, \ref{V1st(Tc)} and  \ref{V1st(QNM)}. 
When this happens hydrodynamical mode is not a propagating one, but 
has some sort of a "diffusive-like" behaviour.


\section{The improved holographic QCD}
\label{ihqcd}

This potential is in a class designed to grasp 
the dynamical features of QCD: the asymptotic freedom
and colour confinement \cite{Gursoy:2008za,Gursoy:2008bu}.
In those aspects it is a more detailed model than the one
used in section \ref{secVQCD}.
Asymptotic freedom is implemented by logarithmic corrections
to the potential in the UV region, while
confinement is detected by a linear dependence of the 
glueballs masses on the consecutive number, i.e., 
$m_n^2\sim n$ for large $n$. This is sometimes
referred to as a linear confinement \cite{Karch:2006pv}. 
The potential we choose has confining IR asymptotic, but
does not include the logarithmic corrections in the UV.


\begin{figure}
	\begin{center}
		\includegraphics[height=.23\textheight]{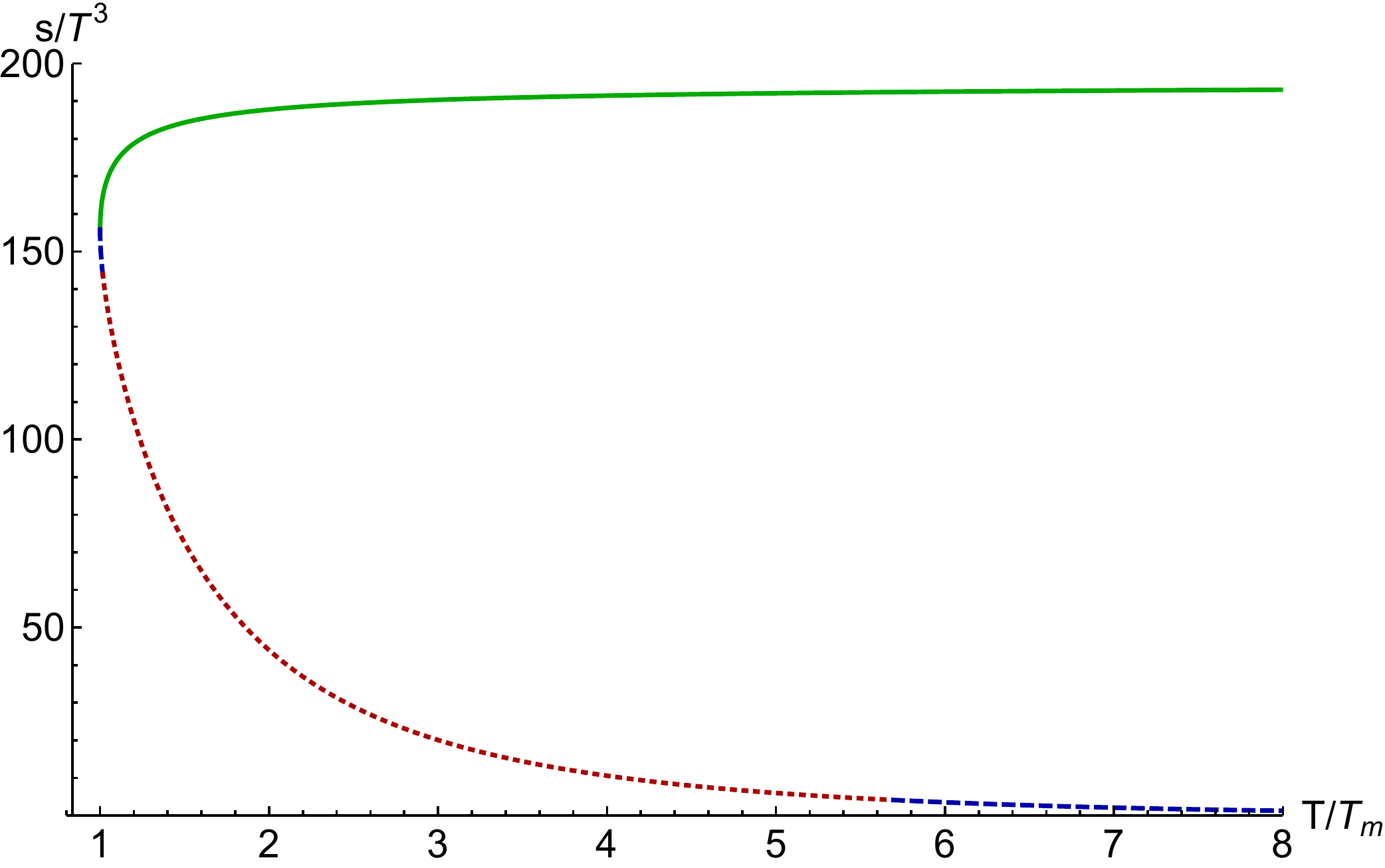}
		\includegraphics[height=.23\textheight]{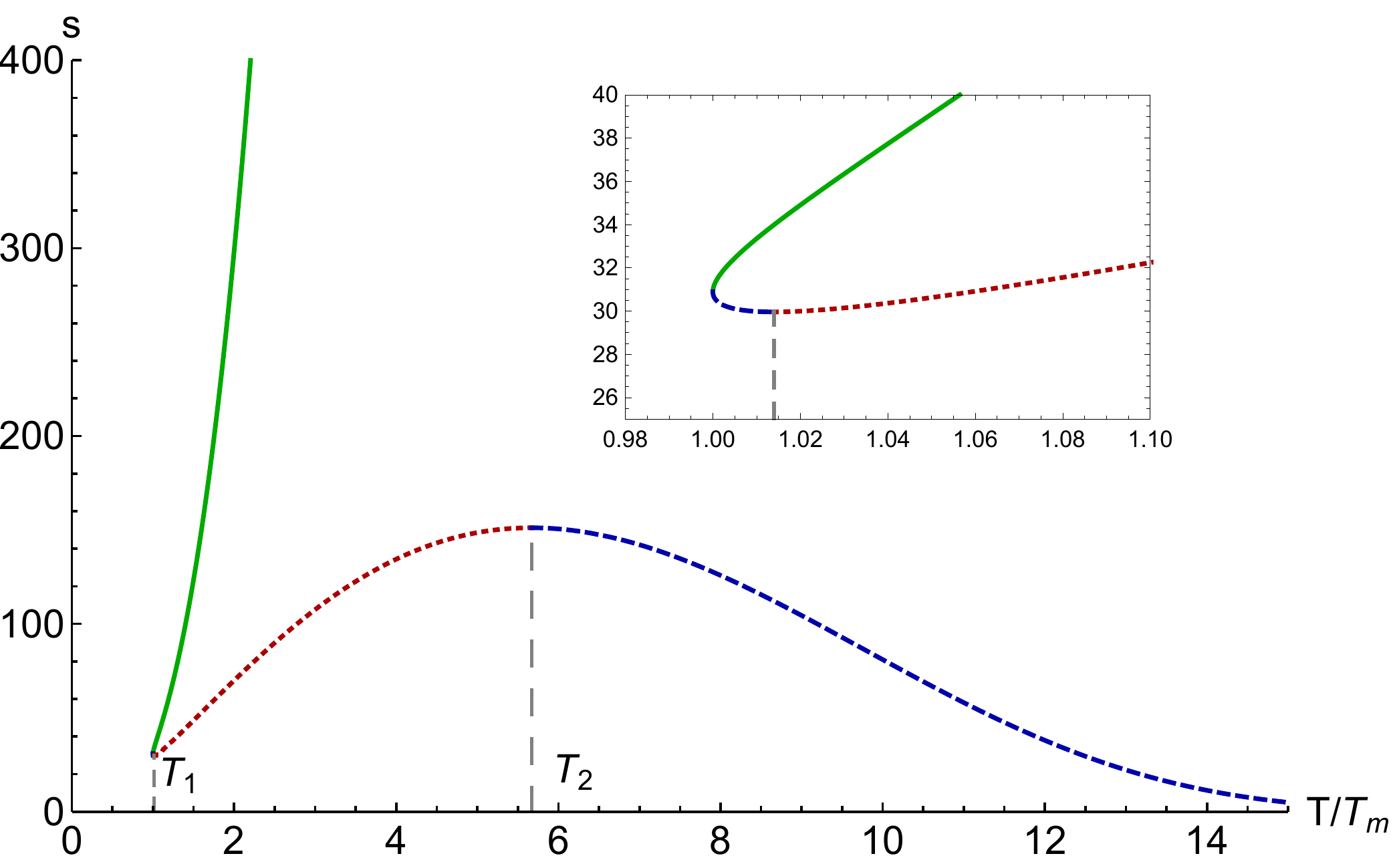}
		\includegraphics[height=.23\textheight]{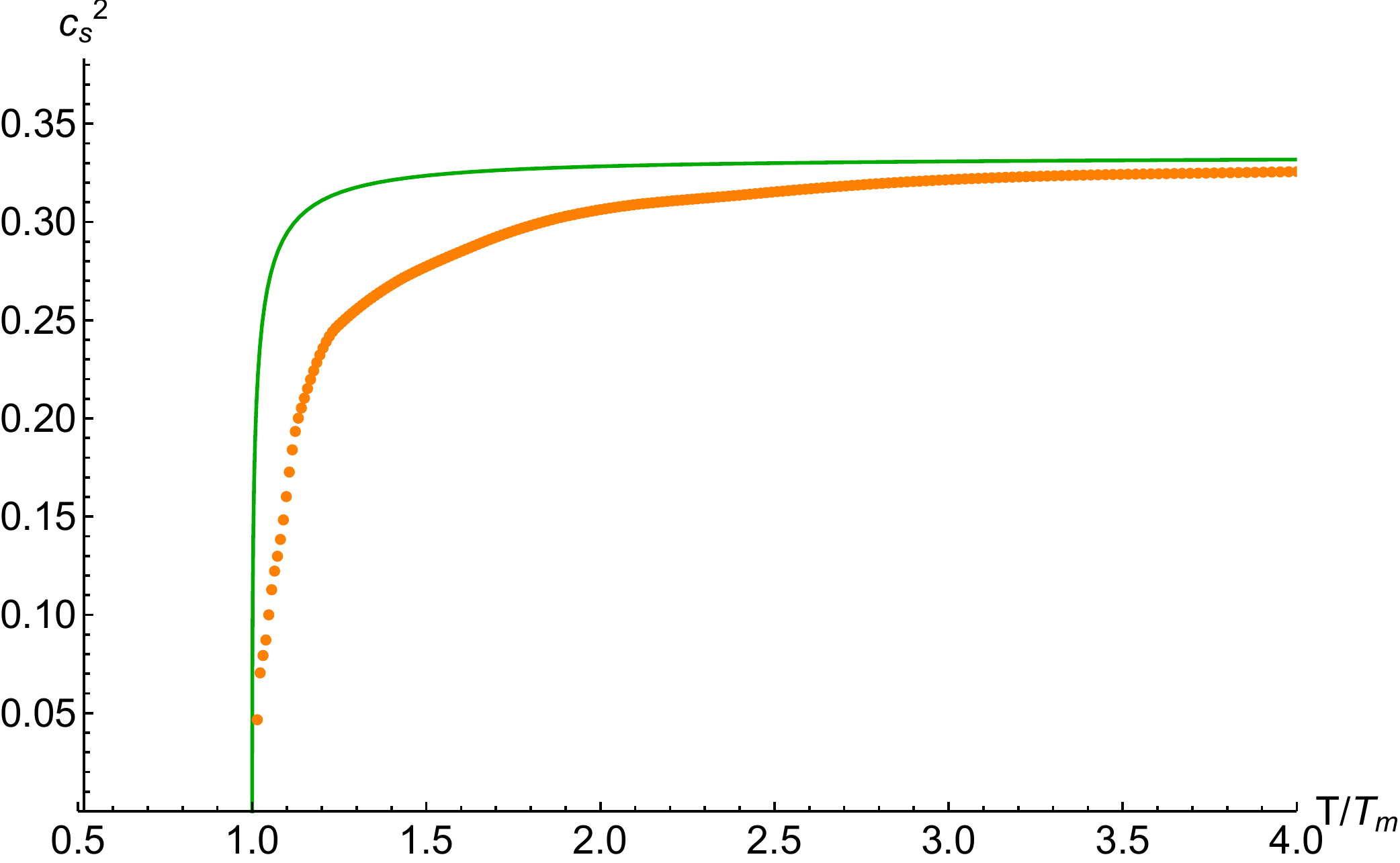}
		\includegraphics[height=.23\textheight]{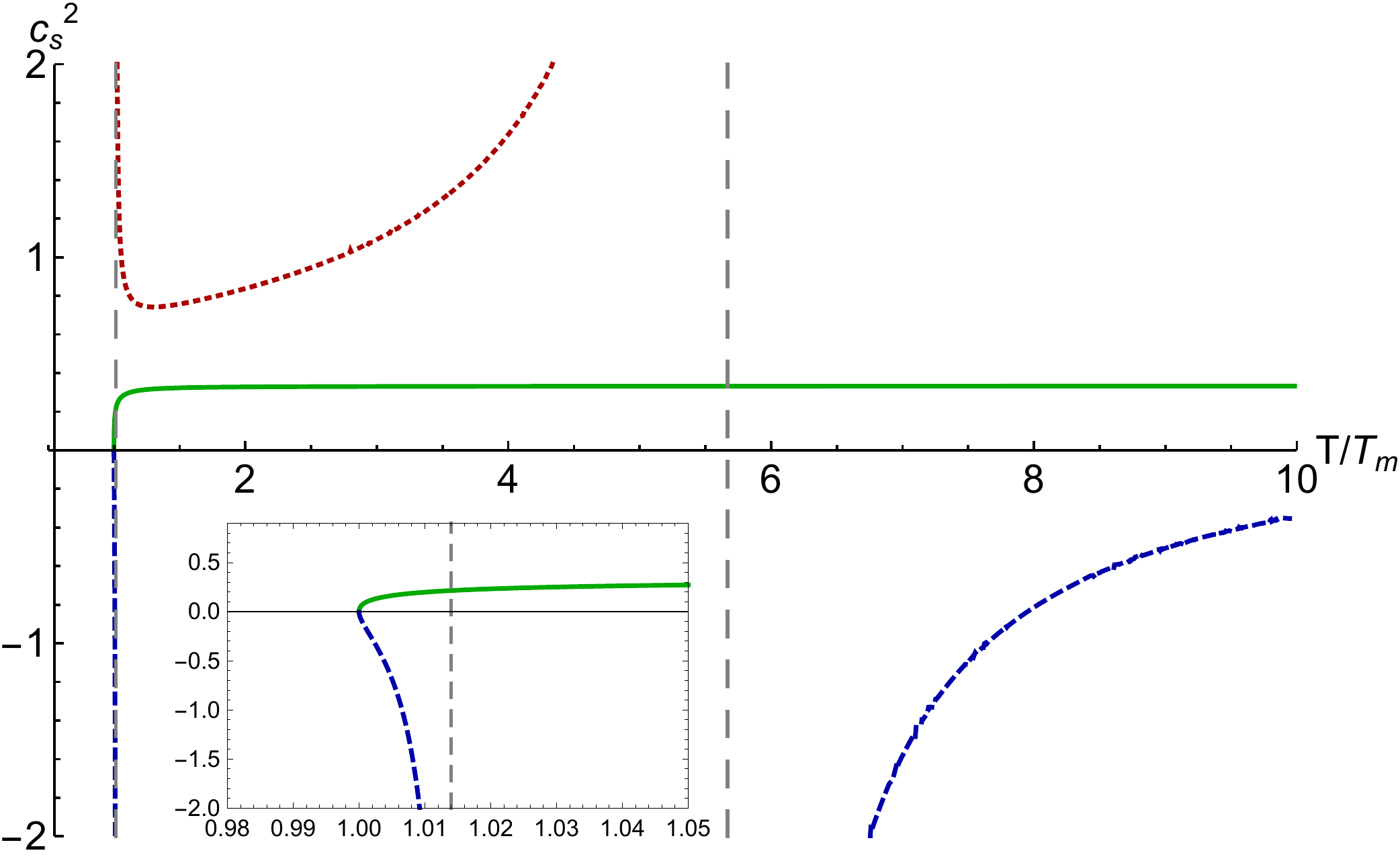}
		\caption{Upper panel: Entropy density as a function of temperature for $V_{\rm IHQCD}$ potential with $\kappa_5=1$.
		Lower panel: Speed of sound squared for the $V_{\rm IHQCD}$ potential (green line), and pure gluon $SU(3)$ 
		lattice data (orange dots) \cite{Boyd:1996bx}. Red and blue dashed lines on the right hand side plot correspond to small
		black hole solutions, which always turn out to be unstable (see text).
		}
		\label{EoS(VIHQCD)}
	\end{center}
\end{figure}


The IHQCD potential determines unique equation of state, with
a rich structure displayed in fig. \ref{EoS(VIHQCD)}.
The two branches of black hole solutions are divided as usual
into large (stable), and small (unstable) configurations. Stable
configurations show behaviour with the usual features characteristic for 
a system with a first order phase transition, and the corresponding
speed of sound is qualitatively similar to the pure glue system \cite{Boyd:1996bx}.
On the contrary, unstable branch consists of two distinct subbranches.
One of them is in a disconnected range of temperatures, $T_m<T<T_1=1.014T_m$ and $T_2=5.67T_m<T$,
and displays spinodal instability signaled 
by the imaginary speed of sound. This in turn implies bubble formation as described
in sec. \ref{1st}. Second sub-branch, $T_1<T<T_2$, shows anomalously
large speed of sound, but does not 
show any instability on the level of equations of state. However, as will be shown below,
in this range of temperatures there exists an unstable non-hydro mode in the QNM spectrum.
  
This system is expected to have a phase transition of a $1^{\rm st}$ order
between a black hole geometry with an event horizon,
and the vacuum confining geometry in the spirit of Hawking-Page phase transition \cite{Hawking:1982dh}.
In principle, to estimate $T_c$ we can find the temperature dependence of the FE along the lines
mentioned in sec. \ref{Background}. In this case non of the methods brings up a decent result.
The direct evaluation of the on-shell
action is corrupted by a numerical instability, while the standard 
thermodynamic relation, $d\mathcal{F}=-s\,dT$, suffers a problem of
correct choice of the reference configuration.
A possible candidate for reference geometry is the one of vanishing
horizon area in the unstable black hole branch. However it has infinite
temperature and it is not clear for us whether it can be used as a proxy
for the thermal gas geometry. Also the unstable branch black holes
exhibit a variety of pathologies (which will be described later)
with increasing $T$. Due to 
the above mentioned difficulties we refrained from estimating the value of $T_c$ in
this particular model. Nevertheless we expect that there exists a critical
temperature $T_c\geq T_m$ where the transition takes place \cite{Gubser:2008ny,Gursoy:2008za}.
This transition changes the geometry substantially.

It is important to note that in this case there exists a minimal temperature $T_m$
below which a black hole solution does not exist. As in the case of $V_{1\rm st}$ the onset 
of instability appears at $T\gtrsim  T_m$ (for configurations with $c_s^2(T)<0$).


\begin{figure}[h!]
	\begin{center}
		\includegraphics[height=.32\textheight]{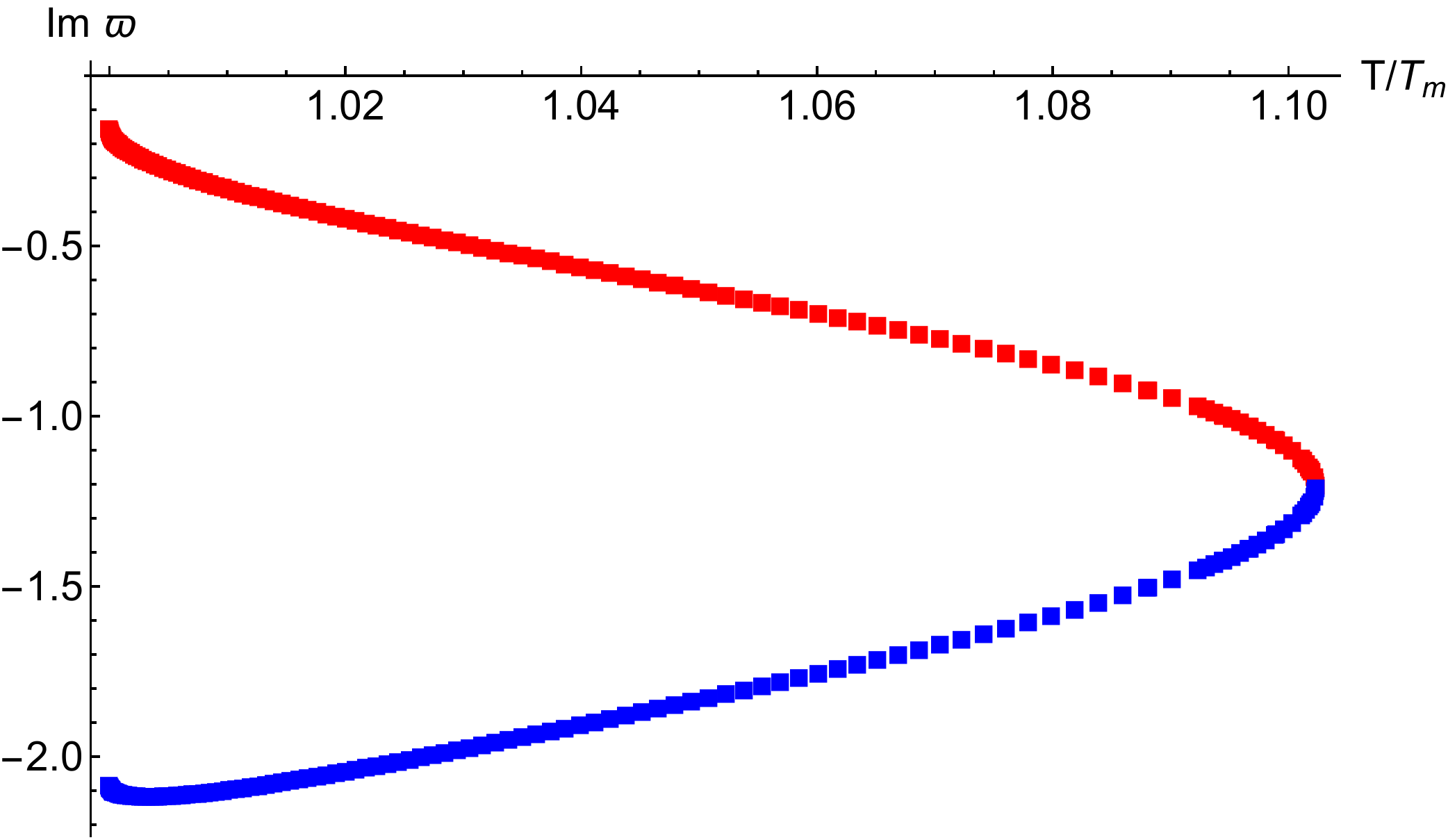}
		\caption{The temperature dependence  of the 
	            non-hydro$_1$ mode at $q=0$. In this temperature range it is spitted   
	            into two purely imaginary branches (red and blue squares).
		}
		\label{VIHQCD(gap)}
	\end{center}
\end{figure}


The different structure of the EoS is reflected in the behaviour of QNM frequencies, which 
indicate the existence of second characteristic temperature $T_{\rm ch}\simeq1.102T_m$. 
The novel effect observed in this system is that for temperatures near the minimal 
temperature the ultralocality property of the first non-hydrodynamic mode is violated.
The mode turns out to be purely imaginary for very low momenta and for temperatures of the range
$T_m\leq T\leq T_{\rm ch}$, and it does not have a structure described in eq. \eqref{OddEven}. 
There are two purely imaginary modes which have the following form
\begin{equation}
\omega_\pm(k) = i \chi(k) \pm i \xi(k)~. 
\end{equation}
In figure \ref{VIHQCD(gap)} we show the temperature dependence of those modes at $k=0$ in the range
where there are purely imaginary. 
As the system is heated further the real part develops, and
the mode becomes the least damped non-hydrodynamic mode of the high-$T$ limit, with the usual structure \eqref{pairRe}. 
It directly comes from the presence of the background scalar field, which breaks the conformal invariance.


\begin{figure}[h!]
	\begin{center}
	\includegraphics[height=.22\textheight]{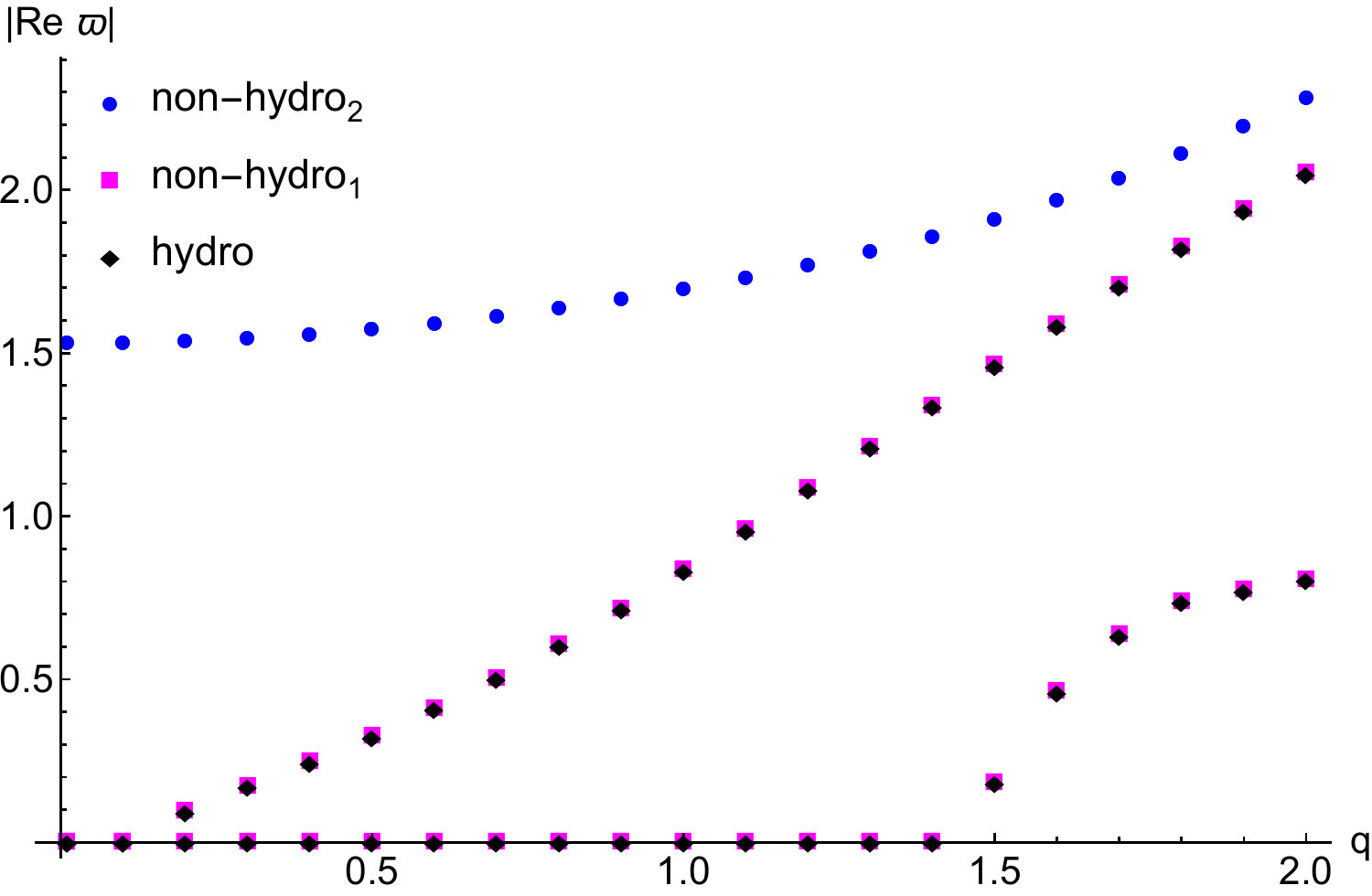}
	\includegraphics[height=.22\textheight]{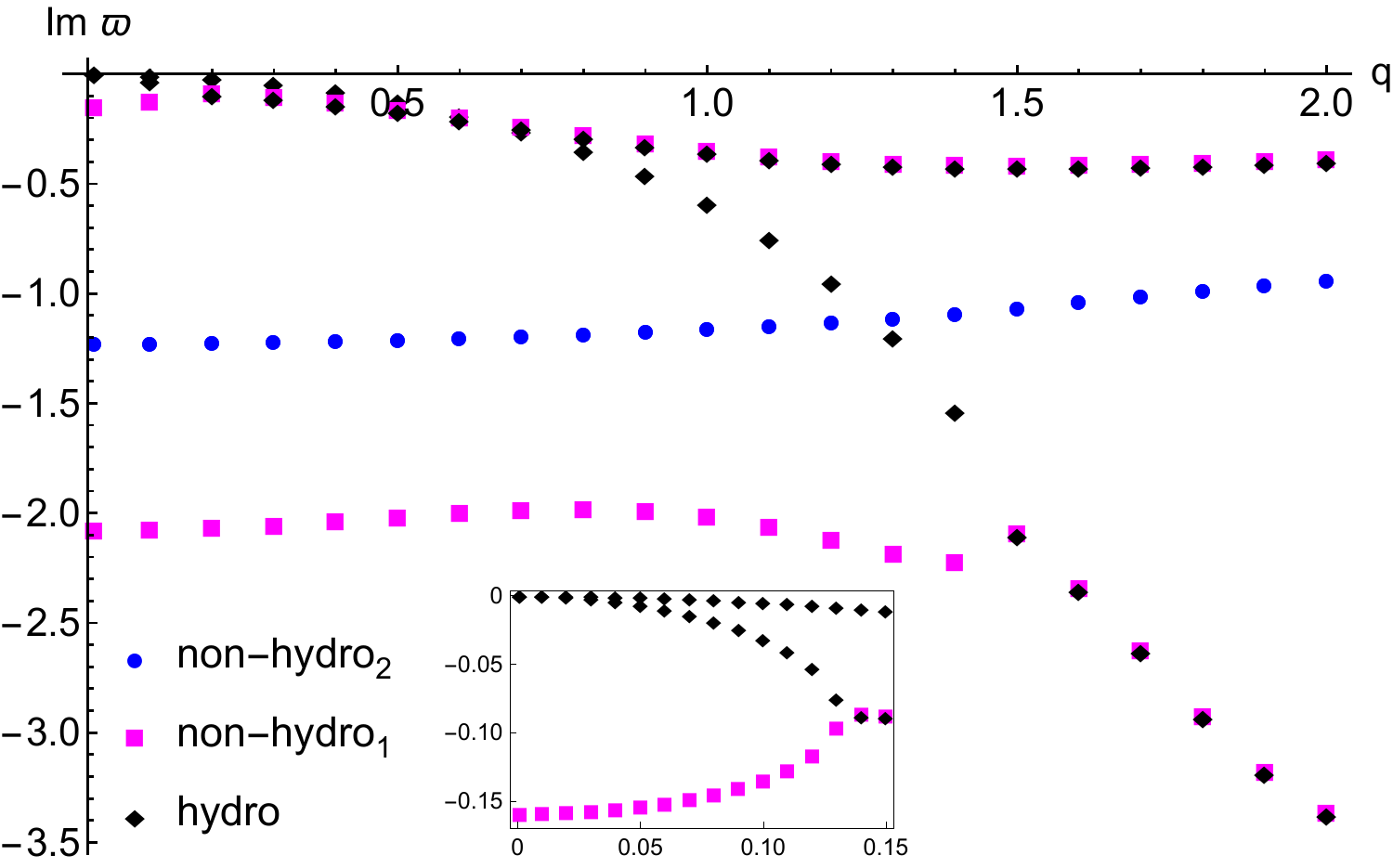}
	\caption{Sound channel quasinormal modes for the potential $V_{\rm IHQCD}$ at  $T=T_m$.
		The non-trivial behaviour of the real part of the modes is shown. In particular in 
		this example the ultralocality of non-hydrodynamic modes is violated, i.e., the non-hydro$_1$ 
		mode attains considerable momentum dependence.
	}
	\label{VIHQCD(QNM)}
\end{center}
\end{figure}


In fig. \ref{VIHQCD(QNM)} we show QNM's in the sound channel
computed for $V_{\rm IHQCD}$ at $T=T_m$. The mode structure 
is different than the one generically present in previous cases. 
First thing which is apparent is that hydrodynamic modes are 
purely imaginary for a range of small momenta.
In addition, there is a small gap between the 
hydrodynamic and non-hydrodynamic degrees of freedom at arbitrary low 
momentum, which in turn implies that the crossing happens at very low value of
$q_c\simeq 0.14$ (see the insert in fig. \ref{VIHQCD(QNM)}). As a matter of fact, in this 
case near the $T_m$ one must always take into account the non-hydrodynamic
degrees of freedom in the description of the system dynamics.
Another absolutely fascinating effect observed exactly at $T_m$ is that the non-hydrodynamic modes, which are purely
imaginary for low momenta,  join with the hydrodynamic modes at some finite momentum $q_J$, and
follow them with increasing $q$. 
This effect is illustrated in fig. \ref{VIHQCD(QNM)}, where the
non-hydro$_{1}$ mode which has two branches joins with the two branches of 
the hydro modes respectively at $q_J\simeq0.14$ and $q_J\simeq1.5$.
In the same time the real part develops with both signs, as expected from general considerations
(see eq. \eqref{pairRe}).
This effect implies the ultralocality violation observed generically 
in other models, and joining does not happen for temperatures higher than the minimal one.
The final observation from fig. \ref{VIHQCD(QNM)} is that the second non-hydrodynamic mode,
referred to as non-hydro$_2$, obeys the ultralocality property, and for high temperatures it
becomes the mode detected in the conformal case \cite{Kovtun:2005ev}.

Interestingly enough a gaped purely imaginary mode was found in \cite{Amado:2009ts}.
System considered there was a holographic dual of superfluidity, and the mode obeying the 
dispersion relation $\omega=-i\gamma(T)-iDk^2$ was found in the superfluid phase.
At the critical temperature $\gamma(T_c)=0$ and the mode becomes an ordinary diffusive mode.
Despite the similarity we have no good physical explanation for this behaviour.


\begin{figure}[h!]
	\begin{center}
		\includegraphics[height=.22\textheight]{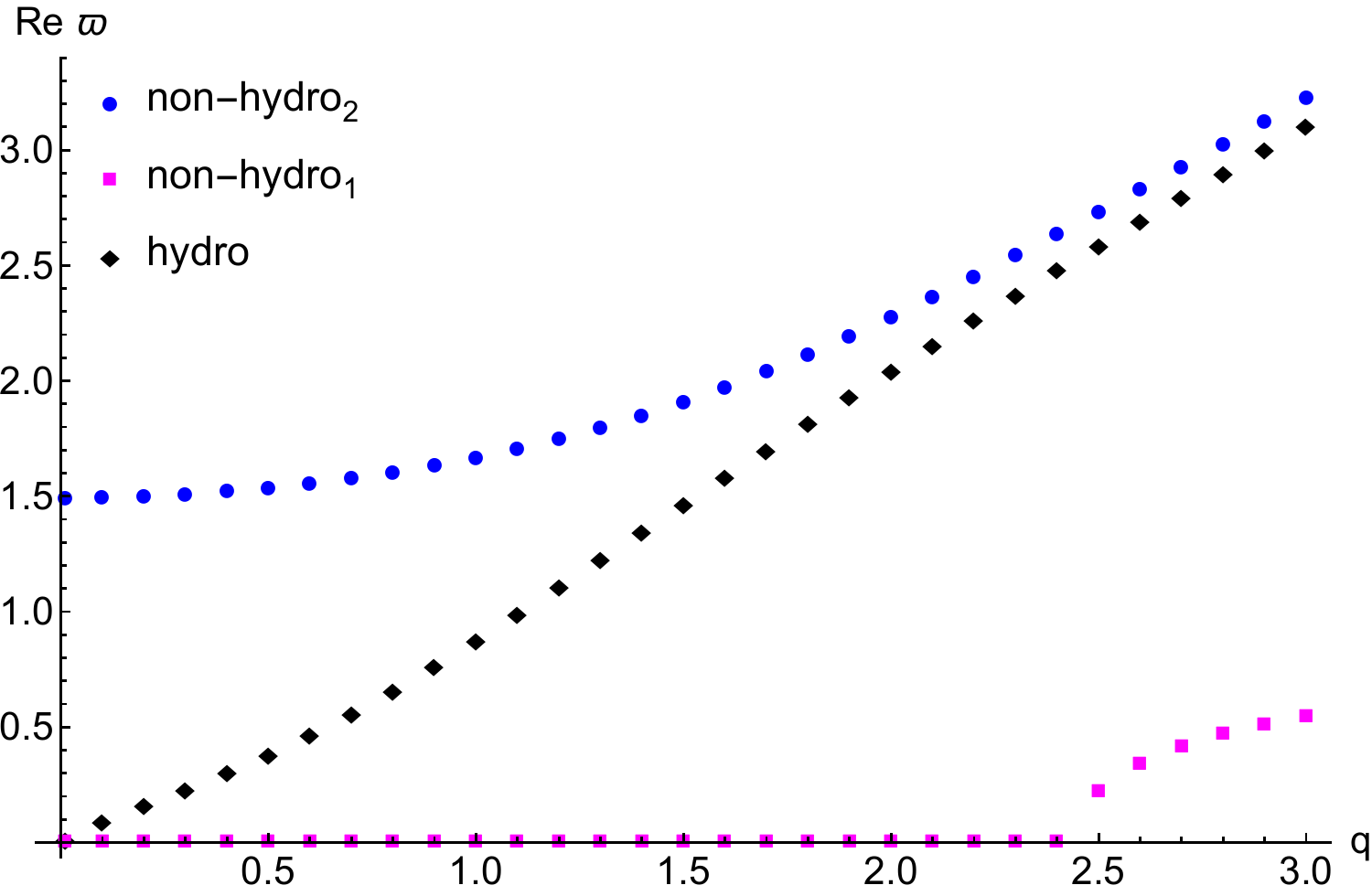}
		\includegraphics[height=.22\textheight]{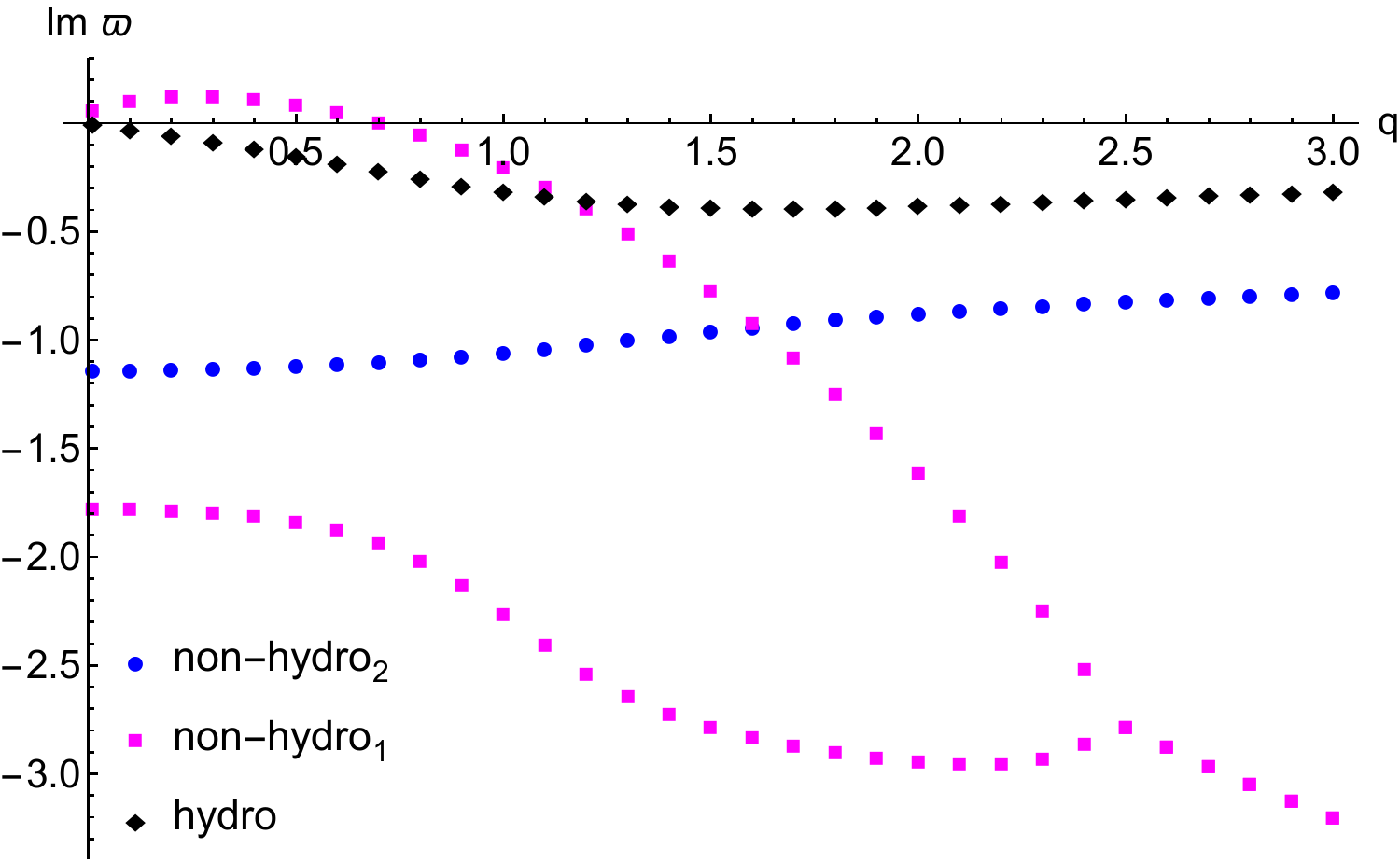}
		\caption{Sound channel quasinormal modes for the potential $V_{\rm IHQCD}$ at $T=1.027T_m$
			in the small black hole branch.
			Real part (left panel) and imaginary part (right panel).
			At this temperatures speed of sound is superluminal and first non-hydro mode
			shows dynamical instability. 
		}
		\label{VQCD(hT-QNM)}
	\end{center}
\end{figure}


The last point to discuss is the spectrum of modes for temperatures, $T_1<T<T_2$, in the small black hole
branch, which shows anomalously large speed of sound. In fact, $c_s^2>1/3$, 
and for some temperatures it is even superluminal, leading 
to causality violation. In this range of temperatures the system does not exhibit 
any instability in thermodynamic quantities.
However, there appears to be a novel {\it dynamical}
instability, signaled by the positive imaginary part of the first non-hydrodynamical mode
\footnote{The nomenclature is chosen because at high temperatures this modes continuously
transforms into first non-hydrodynamic mode.}.
The difference with respect to the usual spinoidal region is that for $k=0$ the mode 
stays positive on the imaginary axis.
Behaviour similar to the one found at $T=T_m$ is also found here: first non-hydro mode stays purely 
imaginary for a range of momenta, and merges with his partner when the real part is developed.
The important difference in this case is that merging is between two modes of the same physical 
nature.


\section{Discussion}
\label{summary}

In the present paper we performed an extensive study of the linearized
dynamics of excitations in strongly coupled field theories in the vicinity
of a nontrivial phase structure of various kinds. 
Generically the effects are visible in the sound channel of the models, while the
shear channel remains less affected. 

We observed a number of novel features which were not present in the conformal case.
For relatively small momenta, the propagating hydrodynamical sound modes become more damped than 
the lowest nonhydrodynamic degrees of freedom. This provides a more stringent restriction on the applicability
of hydrodynamics and indicates the necessity of incorporating these other degrees of freedom 
on appropriate length scales. This is in contrast to the conformal case where a similar phenomenon
only occurred in the shear channel and only at a higher value of momentum.
The richness of phenomena appearing in the linearized regime strongly suggests that it would be important to study the
corresponding real-time dynamics also at the nonlinear level. 

A specific prediction could be anticipated in the potential fitted to lQCD equations of state. 
Qualitative agreement of transport coefficients computed in QCD, and predicted by this model
was known before \cite{Gubser:2008yx} and it is confirmed in our calculations. 
Our novel predictions, however, are concerned with non-hydrodynamic degrees of 
freedom and breakdown of the hydrodynamic description near the QCD critical region.
One specific feature is the ultralocality property obeyed by nonhydrodynamic modes. 
Keeping in mind the qualitative nature of those considerations, it would be 
a very interesting task to compute similar spectrum in lattice QCD.

We study two systems which exhibit different types of the $1^{\rm st}$ order phase transition as 
determined by potentials $V_{1\rm st}$ and $V_{\rm IHQCD}$.
We explicitly determined the instability in the spinoidal branch of both of them,
and for $V_{1\rm st}$ we estimated the length scale for bubble formation.
On top of that, both models posses generic minimal temperature, $T_m$, below which certain
solution cease to exist. From the temperature dependence of the QNM spectrum, in the stable region
of the corresponding EoS, one can 
see the existence of another characteristic temperature, related to the appearance of the diffusive-like
modes, which is slightly higher than $T_m$.

Number of novel phenomena is found in the case of IHQCD potential.
At $T=T_m$ the hydrodynamic and the first non-hydrodynamic modes 
become purely imaginary for low momenta. This implies the violation
of ultralocality property generically observed in other cases \cite{Janik:2015iry,Janik:2015waa}.
One more surprising observation is that instead of the crossing of the 
modes, found generically in the studied models, there is a \emph{``joining''} phenomenon.
From some value of the momentum both the hydrodynamic and first-non hydrodynamic mode
obey the same dispersion relation. In the same time, higher non-hydrodynamic modes admit
ultralocal momentum dependence.

What makes the $V_{\rm IHQCD}$ potential exceptional among the studied cases is a rich 
structure of the small black hole branch solutions. The spectrum of quasinormal modes
shows two type of instabilities. One is the usual spinodal instability, similar to one found
in the $V_{1\rm st}$ case. This appears exactly when the systems shows thermodynamic instability
in equations of state.
Second is an instability triggered by the non-hydrodynamic mode. In this case the relation
between EoS and instabilities is that for configurations which have $c_s^2>1/3$ the first
non-hydrodynamic mode becomes unstable.
Two regions are separated and do not overlap. Up to our knowledge this is the 
first example where such a  dynamical mechanism has been presented.

\vspace{15pt}



{\bf Acknowledgments.} RJ and HS were supported by NCN grant 2012/06/A/ST2/00396, JJ by 
the NCN post-doctoral internship grant DEC-2013/08/S/ST2/00547.
We thank Juergen Engels for providing us with the lattice data for the speed of sound squared 
in the pure gluon sector.
We would like to thank D. Blaschke and P. Witaszczyk for interesting discussions.


\appendix

\section{On-shell action and Free Energy}
\label{AppB}

In this appendix we give some details about asymptotic behaviour of our black hole solutions and we show how, in principle, one can use this expansion to compute the Free Energy. Solving the equations of motion \eqref{eq1}-\eqref{eq4}, close to the boundary one can find the asymptotic form of the general solutions as\footnote{We are interested in the potentials which lead to non-integer conformal dimension $\Delta$, otherwise some additional logarithmic terms may appear in the expansion.}
\bea
&&A(r)=\frac{\ln(r)}{\Delta-4}+\sum_{n,m=0}^{\infty} a_{n\,m} r^{2n+m \Delta/(4-\Delta)}~,\label{asymp1}\\
&&B(r)=\ln\left(\frac{1}{r(4-\Delta)}\right)+\sum_{n,m=0}^{\infty} b_{n\,m} r^{2n+m \Delta/(4-\Delta)}~,\label{asymp2}\\
&&h(r)=1+\sum_{n=0,m=1}^{\infty} h_{n\,m} r^{2n+4 m /(4-\Delta)}~,\label{asymp3}
\eea
such that 
\bea
a_{0\,0}=0~,\qquad b_{0\,0}=0~,\qquad h_{0\, 1}=-c/4,
\eea
where  $c$ is a constant, related to the horizon data,
\be
c=2\, \kappa_5^2\, s\, T~.
\ee
The coefficients $a_{0,m}$ (with $m>1$), $b_{0,m}$ (with $m\geq1$) can be found explicitly in terms of the conformal weight $\Delta$ and the coefficient $a_{0,1}$. By solving the equations of motion order by order, all higher coefficients
$a_{k,m}$, $b_{k,m}$ (with $k>0$) will be fixed in terms of $a_{0,1}$, $\Delta$ and $c$. For a given solution which is unique for a given $r_H=\phi_H$ one can read off numerically $a_{0,1}$ coefficient by studying the near boundary behaviour.

Let us notice that our black hole Ansatz \eqref{linelef} reduces to a thermal gas solution by imposing the $c=0$ condition \cite{Gursoy:2008bu}. In other words, all coefficients in the blackening function $h$ are proportional to positive powers of $c$. 

To compute the Free Energy of a given solution using the holographic renormalization approach one needs to know the boundary counter terms. Introducing the thermal gas solution with a "good" singularity \cite{Gubser:2000nd} as the reference configuration one can calculate the Free Energy
\be
\beta \mathcal{F}=\lim_{\epsilon\rightarrow0}\left(S_{\text{BH}}(\epsilon)-S_{\text{TG}}(\epsilon)\right)~,
\ee 
without having the explicit form of the counter terms, since their contribution in two solutions (black hole and thermal gas) will be canceled. We follow the method which has been explained in detail in appendix C of  \cite{Gursoy:2008bu}. The main difference we want to emphasize is that thanks to the gauge we choose for our radial coordinate, $\phi=r$, we can use the same cut-off on both solutions. 
Computing the On-shell action (including Gibbons-Hawking term) for a given black hole solution and subtracting the corresponding On-shell action of the thermal gas solution one shows that,
\bea
\mathcal{F}=\frac{V_3}{2\kappa_5^2} \, \lim_{\epsilon\rightarrow0}\left\{e^{4A(\epsilon)-B(\epsilon)}\left(6 h(\epsilon) A'(\epsilon)+h'(\epsilon)\right)-6e^{4A(\epsilon)-\tilde{B}(\epsilon)}\sqrt{h(\epsilon)}\tilde{A}'(\epsilon)\right\}~,\label{FE2}
\eea
where $V_3$ is the volume of 3-space and  functions with tilde correspond to the thermal gas solution with the same asymptotic as the black hole. Plugging the near boundary expansion \eqref{asymp1}-\eqref{asymp3} in \eqref{FE2} it is easy to see that the divergent terms will be canceled, namely
\bea
\mathcal{F}&=&\frac{V_3}{2\kappa_5^2}\left(-\frac{c}{4}+\frac{(4-\Delta)^2(2-\Delta)}{2}\left(a_{0 1}-\tilde{a}_{0 1}\right)\right)\nn\\
&=&-\frac{V_3\, s\,T}{4}+\frac{1}{2\kappa_5^2}\frac{(4-\Delta)^2(2-\Delta)}{2}\left(a_{0 1}-\tilde{a}_{0 1}\right)~,
\eea
where $a_{01}$ and $\tilde{a}_{01}$ are the coefficients of the near boundary expansion of the black hole and thermal gas solution respectively. The first term is what we expect for a conformal theory in (3+1)-dimension and the second term corresponds to gluon condensation \cite{Gursoy:2008bu}. 

We would like to note that the subtraction term proportional to $\tilde{a}_{01}$ is constant. One may use this fact to find its value numerically by taking the zero-size limit of the black hole solutions, which corresponds to the "good" singularity definition \cite{Gubser:2000nd}. 

For our V$_{1\rm st} $ potential we find the $a_{01}$ parameter for a given solution by fitting our numerical solution for function $A(r)$ with the asymptotic expansion \eqref{asymp1} up to $\mathcal{O}(r^{40})$. The results are stable and in perfect agreement with the other method explained in section \ref{secThermo}. The corresponding Free Energy is given in the right panel of Fig. \eqref{EoS(V1st)}. Unfortunately, we can use neither this numerical method (because of numerical instability), nor the method using the thermodynamic relation \eqref{eq:Frh} (due to the difficulties sketched in section~7) to compute the Free Energy for the potential V$_{\text{IHQCD}}$.

\section{QNMs equations of motion and numerical details }
\label{app}

In this appendix we show explicitly the QNM equations of motion,
obtained by linearization of the Einstein-Scalar
system of equations for the gauge invariant combinations of the fluctuations.
Using the definition of $Z_1(r)$ and $Z_2(r)$ we can then decouple equations of
motion for sound channel and nonconfomral channel and set them as
\begin{eqnarray}\label{appeq1}
\begin{array}{l}
M_2(r) Z_1''(r)+M_1(r)Z_1'(r)+M_0(r)Z_1(r)+K_0(r)Z_2(r) = 0~,\\
N_2(r) Z_2''(r)+N_1(r)Z_2'(r)+N_0(r)Z_2(r)+L_1(r)Z_1'(r)+L_0(r)Z_1(r) = 0~.
\end{array}
\end{eqnarray}
%
In those equations:
\begin{eqnarray}
	M_2(r) &=& 36 G(r)^2 e^{2 A(r)+B(r)} \left(3 G(r) V'(r)+V(r)\right)\times\nn\\
	&& (k^2 e^{2 B(r)} \left(V(r) \left(6 G'(r)+12 G(r)^2+1\right)+3 G(r) \left(12 G(r)^2+1\right) V'(r)\right)
	+108 \omega ^2 G(r)^2 G'(r))~,\nn
\end{eqnarray}

\begin{eqnarray}
M_1(r) &=&-36 i k^2 \omega  G(r) e^{A(r)+2 B(r)} G'(r) (V(r) (3 G(r)^2 (7-6 G'(r))-6 G'(r)+396 G(r)^4-1)\nn\\
&&+3 G(r) (396 G(r)^4+21 G(r)^2-1) V'(r))+k^2 e^{3 B(r)} (e^{2 A(r)} (12 G(r)^2+1)\nn\\
&& (V(r) (6 G'(r)-24
   G(r)^2+1)+3 G(r) (1-24 G(r)^2) V'(r))^2+108 k^2 G(r)^2 G'(r)\nn \\
   &&(V(r) (6 G'(r)+12 G(r)^2+1)+3 G(r) (12 G(r)^2+1) V'(r))-34992 i \omega ^3 e^{A(r)} G(r)^5 G'(r)^2\nn\\
   &&+11664 k^2 \omega ^2 e^{B(r)} G(r)^4 G'(r)^2~,\nn\\
\nonumber
M_0(r) &=&12 e^{A(r)} G(r) (k^2 e^{A(r)+3 B(r)} (3 G(r) V(r) (12 G(r)^2 (2-9 G'(r))-9 G'(r) (2 G'(r)+1)\\
\nonumber
&&+576 G(r)^4-2) V'(r)+9 G(r)^2 (12G(r)^2+1) (-3 G'(r)+24 G(r)^2-1) V'(r)^2\\
\nonumber
&&+(12 G(r)^2+1) V(r)^2 (-6 G'(r)+24 G(r)^2-1))-972 \omega ^2 G(r)^3 e^{A(r)+B(r)} G'(r)^2V'(r)\\
\nonumber
&&-18 i k^2 \omega  e^{2 B(r)} G(r) G'(r) (V(r) (6 G'(r)+12 G(r)^2+1)+3 G(r) (12 G(r)^2+1) V'(r))\nn\\
&&-1944 i \omega ^3 G(r)^3 G'(r)^2)~,\nn\\
K_0(r) &=& -2 k^2 e^{2 A(r)+3 B(r)} (V(r) (-6 G'(r)+24 G(r)^2-1)+3 G(r) (24 G(r)^2-1) V'(r))\nn\\
&& (k^2 e^{2 B(r)} (V(r) (6 G'(r)+48G(r)^2+1) V'(r)+3 G(r) (12 G(r)^2+1) V'(r)^2+12 G(r) V(r)^2)\nn\\
&&+6 \omega ^2 G(r) (3 G(r) (6 G'(r)+24 G(r)^2-1) V'(r)+(24G(r)^2-1) V(r)))~,\nn\\
N_2(r) &=& G(r) e^{2 A(r)+B(r)} (3 G(r) V'(r)+V(r)) (k^2 e^{2 B(r)} (V(r) (6 G'(r)+12 G(r)^2+1)\nn\\
&&+3 G(r) (12 G(r)^2+1) V'(r))+108 \omega ^2G(r)^2 G'(r))~,\nn\\
N_1(r) &=& -3 e^{A(r)} G(r) G'(r) (e^{A(r)+B(r)} V'(r)+2 i \omega ) (k^2 e^{2 B(r)} (V(r) (6 G'(r)+12 G(r)^2+1)\nn\\
&&+3 G(r) (12 G(r)^2+1)V'(r))+108 \omega ^2 G(r)^2 G'(r))~,\nn\\
N_0(r) &=& G'(r) (-9 i k^2 \omega  G(r)^2 e^{A(r)+2 B(r)} (V(r) (6 G'(r)+12 G(r)^2+1)+3 G(r) (12 G(r)^2+1) V'(r))\nn\\
&&+36 \omega ^2 e^{B(r)} G(r) G'(r)(e^{2 A(r)} (9 G(r)^2 V''(r)+6 G(r) V'(r)+V(r))+9 k^2 G(r)^2)\nn\\
&&+k^2 e^{3 B(r)} (e^{2 A(r)} (V(r) (6 G'(r)+1) V'(r)+3 G(r) (V(r)(6 G'(r)+1) V''(r)+V'(r)^2+4 V(r)^2)\nn\\
&&+108 G(r)^4 V'(r) V''(r)+36 G(r)^3 (V(r) V''(r)+4 V'(r)^2+3 G(r)^2 V'(r) (3 V''(r)+28V(r)))\nn\\
&&+3 k^2 G(r) (V(r) (6 G'(r)+12 G(r)^2+1)+3 G(r) (12 G(r)^2+1) V'(r)))-972 i \omega ^3 e^{A(r)} G(r)^4 G'(r)~,\nn\\
L_1(r)&=&36 G(r) e^{2 A(r)+B(r)} G'(r)^2 \left(3 G(r) V'(r)+V(r)\right)~,\nn\\
L_0(r) &=& 3 e^{A(r)} G'(r)^2 (e^{A(r)+B(r)} (V(r) (-6 G'(r)+24 G(r)^2-1)+3 G(r) (24 G(r)^2-1) V'(r))\nn\\
&&-36 i \omega  G(r) G'(r))~.\nn
\end{eqnarray}
In the $k=0$ case all equations, except for the $Z_2(r)$ which remains coupled to $Z_1(r)$, reduce to scalar field equations.
This case has been studied in ref. \cite{Janik:2015waa}.

In the case of shear channel, $Z_3(r)$, the gauge invariant form is given in \eqref{shearMode} and the QNM equation has the following form,
\bea\label{appeq2}
J_2(r)\, Z_3''(r)+J_1(r)\,Z_3'(r)+J_0(r)\,Z_3(r)=0~,
\eea
where
\bea
J_2(r)&=&2 e^{2 A(r)+B(r)} \left(3 A'(r) V'(r)+V(r)\right) \left(\omega ^2 \left(6 A'(r) B'(r)-1\right)+2 k^2 e^{2 B(r)} \left(3 A'(r) V'(r)+V(r)\right)\right)~,\nn\\
J_1(r)&=&e^{A(r)} \big(-4 i k^2 \omega  e^{2 B(r)} \left(6 A'(r) B'(r)-1\right) \left(3 A'(r) V'(r)+V(r)\right)+4 k^2 e^{A(r)+3 B(r)} \left(4 A'(r)-B'(r)\right)\nn\\
&&\left(3 A'(r) V'(r)+V(r)\right)^2-\omega ^2 e^{A(r)+B(r)} V'(r) \left(1-6 A'(r) B'(r)\right)^2-2 i \omega ^3 \left(1-6 A'(r) B'(r)\right)^2\big)~,\nn\\
J_0(r)&=&\left(6 A'(r) B'(r)-1\right) \big(-i \omega  e^{A(r)} \big(k^2 e^{2 B(r)} \left(2 V(r) \left(7 A'(r)-B'(r)\right)+\left(42 A'(r)^2-1\right) V'(r)\right)\nn\\
&&+3 \omega ^2 A'(r) \left(6 A'(r) B'(r)-1\right)\big)+6 k^2 \omega ^2 e^{B(r)} A'(r) B'(r)+2 k^4 e^{3 B(r)} \left(3 A'(r) V'(r)+V(r)\right)\nn\\
&&-k^2 \omega ^2 e^{B(r)}\big)~.\nn
\eea
At $k=0$ this equations reduces to a equation of motion of minimally coupled massless scalar field
which was studied in \cite{Janik:2015waa}.

In order to solve equations \eqref{appeq1} and \eqref{appeq2}, 
which are linear ordinary differential equations, we use 
spectral discretization with Chebyshev polynomials \cite{Grandclement:2007sb}.
The resulting matrix equation is of polynomial character in the mode 
frequency $\omega$ and we determine QNMs by evaluating the 
determinant of the matrix and setting it to zero. Corresponding vectors in the kernel
of this matrix are discretized versions of the radial dependence of the desired solutions.
To find physical solutions with the right boundary behaviour we fit the tail of the function
to a form obtained from small $r$ (near the conformal boundary) analysis. In particular for the modes $Z_1(r)$ and $Z_2(r)$
this procedure is non-trivial and the asymptotic expansions are given in eq. \eqref{form}. 


\end{document}